\documentclass[fleqn,usenatbib]{mnras}

\usepackage{newtxtext,newtxmath}

\usepackage[T1]{fontenc}

\DeclareRobustCommand{\VAN}[3]{#2}
\let\VANthebibliography\thebibliography
\def\thebibliography{\DeclareRobustCommand{\VAN}[3]{##3}\VANthebibliography}

\usepackage{graphicx}	
\usepackage{amsmath}	
\usepackage{multicol}   

\title[LGS scattering and Raman emission GTC measurements]{Laser Guide Star uplink beam: scattering and Raman emission measurements with the 10.4m Gran Telescopio CANARIAS}

\author[G. Lombardi et al.]{
G. Lombardi,$^{1,2}$\thanks{E-mail: gianluca.lombardi@gtc.iac.es}
D. Bonaccini Calia,$^{3}$
M. Centrone,$^{4}$
A. de Ugarte Postigo$^{5}$
and S. Geier$^{1,2}$
\\
$^{1}$Gran Telescopio CANARIAS S.A., c/ Cuesta de San Jos\'e s/n, E-38712 Bre\~{n}a Baja, La Palma, Spain\\
$^{2}$Instituto de Astrof\'{\i}sica de Canarias, V\'ia L\'actea s/n, E-38205 La Laguna, Tenerife, Spain\\
$^{3}$European Southern Observatory, Karl-Schwarzschild-Str. 2, D-85748 Garching bei Muenchen, Germany\\
$^{4}$INAF-OAR National Institute for Astrophysics, Via Frascati 33, I-00078 Monte Porzio Catone, Roma, Italy\\
$^{5}$Artemis, Observatoire de la C\^ote d’Azur, Universit\'e C\^ote d’Azur, Boulevard de l’Observatoire, FR-06304 Nice, France
}

\date{Accepted 2022 August 01. Received 2022 August 01; in original form 2021 October 27}

\pubyear{2022}

\begin{document}
\label{firstpage}
\pagerange{\pageref{firstpage}--\pageref{lastpage}}
\maketitle

\begin{abstract}
Laser Guide Star Adaptive Optics (LGS-AO) is becoming routine in several astronomical observatories. The use of powerful lasers generates sensible Raman emissions on the uplink laser beam path, plus secondary Rayleigh scattering from atmospheric molecules and Mie scattering from aerosols. This paper reports the results of a campaign done with the 10.4m Gran Telescopio CANARIAS (GTC); this campaign was undertaken to assess the spectral and photometric contamination coming from a 589 nm laser uplink beam scattering and Raman emission induced on the GTC spectro-imager OSIRIS by laser launched $\sim$1 km off-axis. The photometric contamination is due to primary and secondary scattering of the uplink photons, as well by the Raman inelastic scattering. We have propagated the laser beam creating a mesospheric LGS, then pointed and focused the GTC telescope toward the uplink laser beam, at different heights and up to the LGS, taking into account the observing geometry. In our observations, the Raman emissions for O$_{2}$ and N$_{2}$ vibrational lines are visible at 20 km, weakening with altitude and becoming undetectable above 30 km. The scattering of the focused uplink beam is detectable at less than $\pm$0.2 arcmin from the center of the beam, while for the focused LGS the scattering is narrower, being detectable at less than $\pm$0.1 arcmin around the plume. Recommendations for Laser Traffic Control Systems (LTCS) are given accordingly.
\end{abstract}



\begin{figure*}
   \centering
   \includegraphics[width=1.00\textwidth]{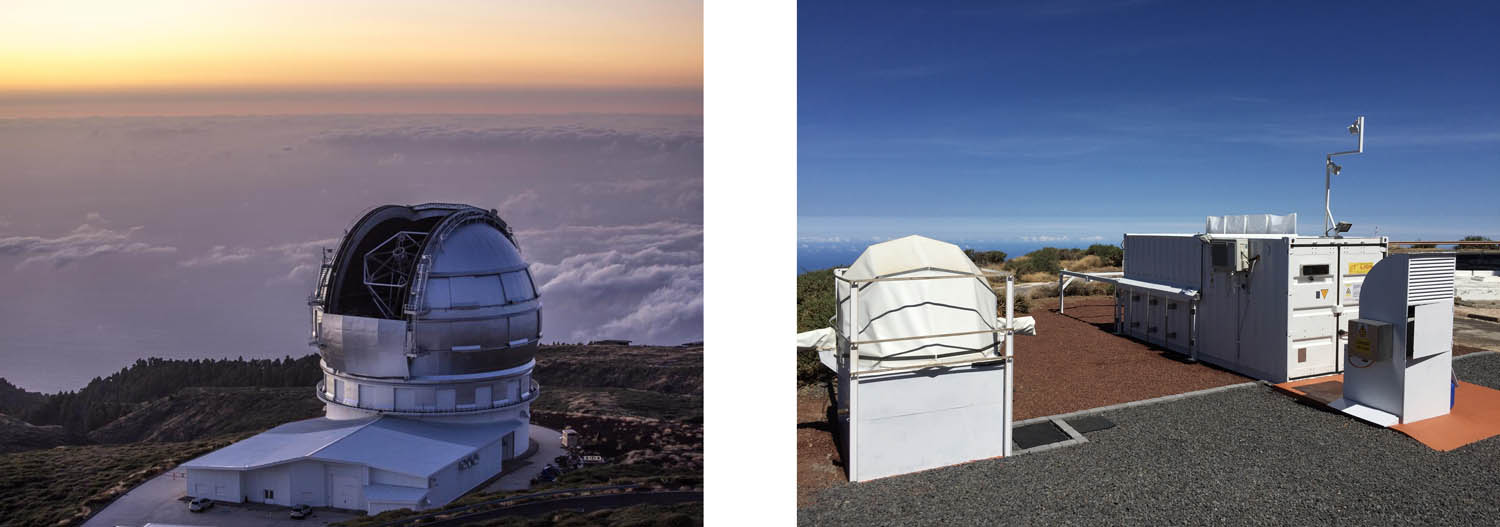}
   \caption{(left) The 10.4m Gran Telescopio CANARIAS at the ORM (Credit: Daniel L\'opez/IAC). (right) WLGSU configuration at ORM next to the WHT, with a small receiver telescope dome (on the left) and the sliding-roof container accommodating the laser beam transmitter. The small receivers of the WLGSU were not used in the experiment, the 10.4m GTC telescope was used as receiver instead.}
   \label{fig:WLGSUatORM}
\end{figure*}
\section{Introduction}
Laser Guide Star (LGS) systems are used to create an artificial star in the sky. In modern Astronomy, Laser Guide Stars are used to support LGS Adaptive Optics (LGS-AO) technologies. Two types of LGS are available for such purpose, sodium and Rayleigh beacon guide stars. Nevertheless, in Astronomy sodium LGS are actually the type used in the vast majority of cases. Sodium LGS are created by the excitation of mesospheric sodium molecules at 589 nm located at about 90 km above sea level, distributed among the mesosphere for several kilometers.\\
In recent years, a transportable and compact Laser Guide Star Unit system was built by the European Southern Observatory (ESO), in which an experimental 20W Continuous Wavelength (CW) laser at 589 nm, based on the ESO patented Narrow Band Fiber Raman Amplifier technology (Patent no. 8773752), has the laser head attached directly to a 0.38 m refractive launch telescope (\citealt{bonac10}). The so-called \textit{Wendelstein Laser Guide Star Unit} (WLGSU) has been designed, assembled, tested in the ESO laboratory and successfully commissioned in the summer of 2011, in Bavaria, Germany (\citealt{bonac11}). The goal of the WLGSU is to create a facility for field test on Laser Guide Star systems technologies, while also supporting strategic experiments related to LGS-AO technologies.\\
The photometry and the emission spectrum measured on the laser uplink beam are important for the operations aspects of the astronomical observatories with several telescopes on site. Software tools (\citealt{summ03}; \citealt{schal10}; \citealt{gaug18}) are used to avoid that a telescope Field of View (FoV) at the observatory crosses or gets too close to the laser uplink beam while pointing and tracking a science target. Furthermore, it is important to avoid that photometric and spectral contamination from the scattering in the vicinity of the laser beam disturbs the observations. At which distance from the laser beam should the software prevent observations/laser propagation? The measurement of the scattering from the uplink beam is thus needed.\\
\citet{vogt17} given a detailed study of Raman emissions detected by MUSE  at Very Large Telescope (VLT) UT4. The study concerned self-contamination of the same telescope at 25 km above ground only. \citet{vogt18} confirmed that Raman lines scattering is also linked to the cleanliness of the telescope mirrors surface.\\
In order to investigate the emission spectrum and the photometry along the uplink laser beam at different heights, including the LGS itself, four observing runs, each of four hours, have been granted under Program GTC3-16ITP using the OSIRIS spectro-imager (\citealt{cepa00} and \citealt{cepa10}) at the 10.4m Gran Telescopio CANARIAS (GTC) and the WLGSU at the Observatorio del Roque de Los Muchachos (ORM) in the island of La Palma (Canary Islands, Spain). The first two runs on 16 and 17 September 2016 were useful to implement an observing strategy, while on 6 and 7 July 2017 the strategy was finally tested and science observations were performed.\\
The use of a 589 nm laser for LGS at the Canary Islands is ongoing since January 2015, when the ESO WLGSU has been installed in Tenerife at the Observatorio del Teide to carry on return flux experiments (\citealt{bonac16}; \citealt{holz16} and \citeyear{holz16a}). Since July 2016, the WLGSU has been installed at the ORM next to the William Herschel Telescope (WHT), to perform different LGS-AO field experiments and studies related to Extremely Large Telescopes (ELT) and 8-10m class telescopes with the collaboration of the Durham University, the Observatoire de Paris Meudon-LESIA and the German Aerospace Center (DLR). The main goals of the experiments are:
\begin{description}
    \item[$-$] field validating the LGS-AO wavefront sensing, when the LGS image is elongated on the wavefront sensor, as it will be for future ELTs and the GTC. This is achieved using the WHT CANARY AO system which is funded by the EU as pathfinder for the European ELT AO (\citealt{bonac02}, \citeyear{bonac10}, \citeyear{bonac11}, \citeyear{bonac12},; \citealt{grat12}; \citealt{bas17} and \citeyear{bas18});
             \vspace{0.1cm}

\item[$-$] test AO operations with elongated LGS in the E-ELT geometry configuration (\citealt{bonac11} and \citeyear{bonac12}; \citealt{rous15});
         \vspace{0.1cm}

\item[$-$] the feasibility of line-of-sight sodium profile measurements via partial CW laser modulation (\citealt{cas17});
         \vspace{0.1cm}

\item[$-$] the use of LGS to mitigate the effects of the atmospheric turbulence in ground-to-space optical communications (\citealt{mata17}; \citealt{bonac21});
         \vspace{0.1cm}

\item[$-$] test the use of LGS in daytime AO for Solar observations (\citealt{alaluf21});
         \vspace{0.1cm}

\item[$-$] study the photometric and spectral contamination from the scattering near the laser beam along the uplink at different heights, including the LGS itself, in the scenario of multi-laser operations at crowded observatories (\citealt{lom17} and this Paper).
\end{description}
With this Paper we report final results from the Program scientific observations performed on the nights of 6 and 7 July 2017 concerning the photometric and spectroscopic measurements and the observed Raman emission of atmospheric molecules along the uplink beam.\\
\begin{figure*}
   \centering
   \includegraphics[width=1.00\textwidth]{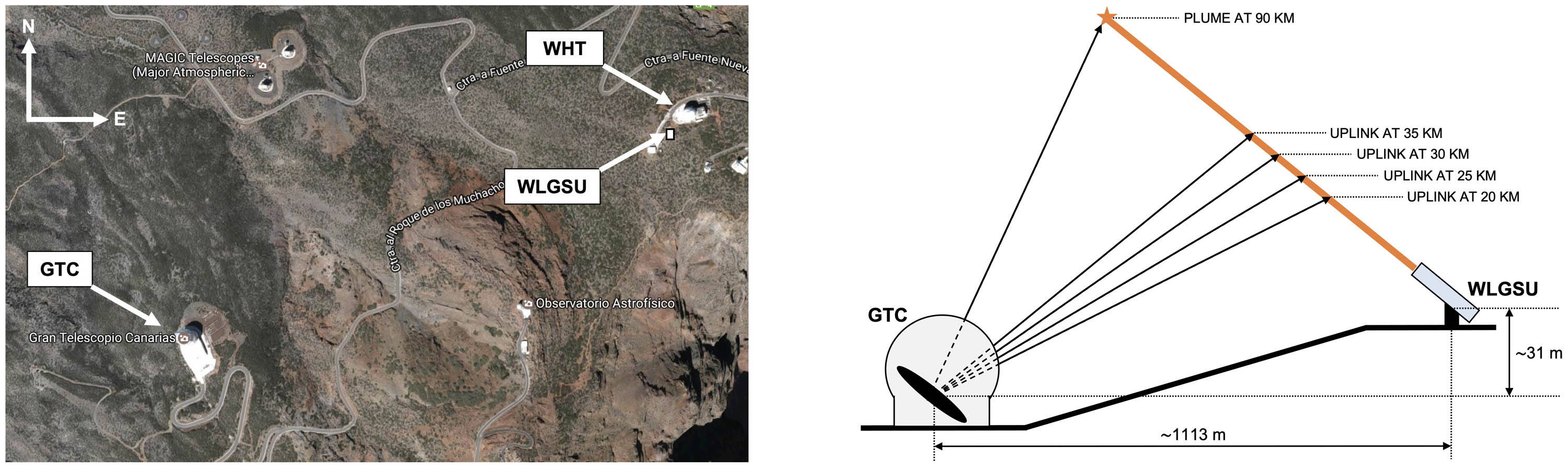}
    \vspace{-0.5cm}
   \caption{(left) Relative positions of the GTC and the WLGSU at ORM (Google Maps cortesy). (right) Schematic representation of the strategy used for the observations of the laser uplink beam and the LGS with the GTC (see Section \ref{sub:ObsStrat}).}
              \label{fig:FigStrategy}
\end{figure*}
       
\vspace{-0.4cm}
\section{Setup, observing strategy and data reduction}
\subsection{Instrument setup}
The setup to conduct the experiment was composed of two parts (see Figure \ref{fig:WLGSUatORM}): a remotely operated 589 nm 20W CW laser transmitter (Tx) located at the WLGSU container, next to the William Herschel Telescope; the receiver telescope was the GTC, equipped with the OSIRIS spectro-imager mounted at the Nasmyth B of the 10.4m telescope. 

\subsection{WLGSU setup}
The laser of the WLGSU transmitter unit is based on a narrow-band Raman Fiber amplifier at 1178 nm and a second harmonic generation unit (\citealt{Taylor09}). It offers wavelength, polarization and amplitude modulation controls. The laser emission is composed of two single frequency lines, each with an emission linewidth of 1.6 MHz, linearly polarized. Most of the emission is tuned to the sodium D$_{2a}$ line at the vacuum wavelength of 589.1591 nm. The second emission line is used for the repumping. It has 12\% of the laser power, tuned at the D$_{2b}$ frequency sideband (1.713 GHz separation). The laser launch telescope emits circularly polarized light in order to maximize the efficiency of optical pumping of the sodium D$_{2}$ transitions (\citealt{holz10}). The Raman amplifier and the second harmonic generation unit form together the assembly called "Wendelstein laser head". 
The Wendelstein laser head is mounted directly on a carbon-fiber-based, 0.38 m diameter refractive launch telescope, installed on an ASTELCO Systems NTM-500 direct-drive motorized mount, anchored on a fixed steel pier mechanically isolated from the container. The emitted laser beam is gaussian, 300 mm in diameter at the 2\% intensity level, 210 mm in diameter at the 13.5\% intensity level.\\
The laser Tx unit was controlled remotely from the GTC control room. The WLGSU implements the standard and local safety measures related to the use of Class IV lasers at the ORM. One of the authors was nonetheless in the Tx area, as direct observer for safety reasons. A Laser Traffic Control System (LTCS) is implemented at ORM to prevent the laser uplink beam to collide with the line-of-sight of the other telescopes, during the experiments.\\

\begin{table}
 \small
     \caption[]{Coordinates of the GTC (telescope pupil) and the WLGSU.}
         \label{TabORM}
		\begin{center}
         \begin{tabular}{r c c c}
            \hline
                  &  \textbf{Longitude}  & \textbf{Latitude}  & \textbf{Elevation}\\
            \hline
            \textbf{GTC} & $17^\circ 53' 30''.8$W & $28^\circ 45' 23''.8$N & 2300 m.a.s.l. \\
            \noalign{\smallskip}
            \textbf{WLGSU} & $17^\circ 52' 54''.1$W & $28^\circ 45' 38''.1$N & 2331 m.a.s.l. \\
            \hline
         \end{tabular}
         \end{center}
\end{table}
\begin{table*}
\small
\caption{Summary of performed observations on the night of the 7th of July 2017. Photometric standard star SA95-193 (\citealt{lan92}) was observed for the calibration of the Broad Band images, while the spectra flux calibration was performed using the BD+33-2642 (\citealt{oke90}) spectrophotometric standard star.}
\begin{center}
\begin{tabular}{crcccccccccccc}
\hline
            \noalign{\smallskip}
 & & & & \multicolumn{3}{c}{\textbf{BROAD BAND IMAGING}} & & \multicolumn{3}{c}{\textbf{LONG SLIT SPECTROSCOPY}} & & \\
\noalign{\smallskip}
            \cline{5-7}
            \cline{9-11}
\noalign{\smallskip}
            \noalign{\smallskip}

\textbf{Altitude} &  \multicolumn{1}{c}{\textbf{Target}} & \textbf{Laser} & & \textbf{Filter} & \textbf{Binning} & \textbf{Exp.Time} & & \textbf{VPH + Slit} & \textbf{Binning} &  \textbf{Exp.Time} & & \textbf{M2 defocus} & \textbf{Seeing}\\
            \noalign{\smallskip}
\hline
            \noalign{\smallskip}
\textbf{90 km} &  \textsc{Plume} &  ON & & Sloan R & 2 $\times$ 2 &  4 sec & & R2500R + s0.6'' & 1 $\times$ 1 &  40 sec & & \textbf{+1.9 mm} & 0.8''\\
               &    \textsc{Sky} & OFF & & Sloan R & 2 $\times$ 2 &  4 sec & & R2500R + s0.6'' & 1 $\times$ 1 &  40 sec & & \\
            \noalign{\smallskip}
\hline
            \noalign{\smallskip}
\textbf{35 km} &  \textsc{Uplink} &  ON & & Sloan R & 2 $\times$ 2 & 30 sec & & R2500R + s0.6'' & 1 $\times$ 1 & 400 sec & & \textbf{+3.0 mm} & 0.8''\\
               &    \textsc{Sky} & OFF & & Sloan R & 2 $\times$ 2 & 30 sec & & R2500R + s0.6'' & 1 $\times$ 1 & 400 sec & & \\
            \noalign{\smallskip}
\hline
            \noalign{\smallskip}
\textbf{30 km} &  \textsc{Uplink} &  ON & & Sloan R & 2 $\times$ 2 & 20 sec & & R2500R + s0.6'' & 1 $\times$ 1 & 300 sec & & \textbf{+4.0 mm} & 0.6''\\
               &    \textsc{Sky} & OFF & & Sloan R & 2 $\times$ 2 & 20 sec & & R2500R + s0.6'' & 1 $\times$ 1 & 300 sec & & \\
            \noalign{\smallskip}
\hline
            \noalign{\smallskip}
\textbf{25 km} &  \textsc{Uplink} &  ON & & Sloan R & 2 $\times$ 2 & 20 sec & & R2500R + s0.6'' & 1 $\times$ 1 & 150 sec & & \textbf{+4.5 mm} & 1.0''\\
               &   \textsc{Sky} & OFF & & Sloan R & 2 $\times$ 2 & 20 sec & & R2500R + s0.6'' & 1 $\times$ 1 & 150 sec & & \\
            \noalign{\smallskip}
\hline
            \noalign{\smallskip}
\textbf{20 km} &   \textsc{Uplink} &  ON & & Sloan R & 2 $\times$ 2 &  8 sec & & R2500R + s0.6'' & 1 $\times$ 1 & 150 sec & & \textbf{+6.0 mm} & 0.6''\\
               &    \textsc{Sky} & OFF & & Sloan R & 2 $\times$ 2 &  8 sec & & R2500R + s0.6'' & 1 $\times$ 1 & 150 sec & & \\
            \noalign{\smallskip}
\hline
\end{tabular}
\end{center}
\label{tab:ObsStrat}
\end{table*}

\subsection{OSIRIS instrument setup}\label{sub:OSIRISsetup}
OSIRIS is a spectro-imager for the optical wavelength range, that at the time of the observations was located in the Nasmith B focus of GTC. In the present the instrument is installed at the Cassegrain focal station of the telescope.  OSIRIS provides standard Broad Band imaging and Long Slit Spectroscopy, and additional capability such as the Narrow Band Tunable Filters imaging, charge-shuffling and Multi Object Spectroscopy. OSIRIS covers the wavelength range from 0.36 to 1.05 $\mu$m with a total FoV of $7.8 \times 8.5$ arcmin ($7.8 \times 7.8$ arcmin unvignetted), and $7.5 \times 6.8$ arcmin, for direct imaging and MOS respectively. For this experiment, we have used OSIRIS in two instrument modes: Broad Band imaging and Long Slit Spectroscopy.\\
Broad Band imaging was performed to obtain the photometry of the laser uplink and to measure the contamination introduced by the beam scattering. A Sloan R filter in standard $2 \times 2$ binning (0.254 arcsec/px) was used. The choice of the filter was driven by its wavelength coverage together with the best achievable sensitivity at 589 nm (90.4\%). The detector binning allowed a 4 times shorter exposure time to obtain enough counts from the uplink beam for a reliable photometric analysis at every altitude.\\
Long Slit Spectroscopy let us investigate the spectrum emission from the laser uplink beam and the LGS plume. Our goal was to get the OSIRIS highest achievable resolution ($R$) with the proper wavelength coverage ($\Delta\lambda$). For that the R2500R volume-phased holographic grating  ($R = 2475$; $\Delta\lambda = 5575-7685$ \r{A}) coupled with the 0.6 arcsec slit and $1 \times 1$ binning (0.508 \r{A}/px) were set.

\subsection{Observing strategy}\label{sub:ObsStrat}
\begin{figure}
   \centering
   \includegraphics[width=0.48\textwidth]{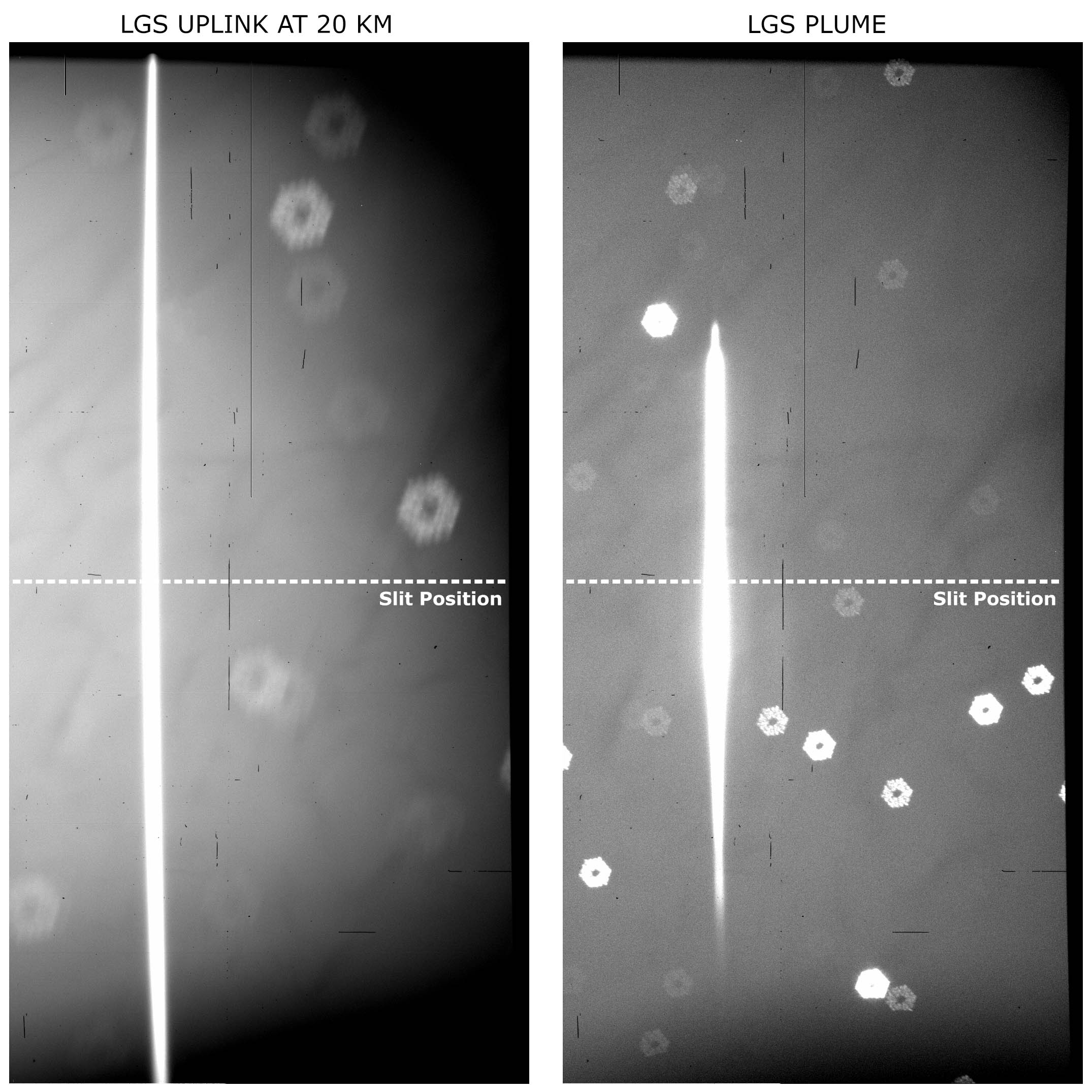}
   \caption{Examples of OSIRIS CCD2 Broad Band raw images of the uplink beam at 20 km (left) and the laser plume (right) with the Sloan R filter. The figure reveals strongly defocused field stars crossing the OSIRIS FoV, uncovering the GTC segmented pupil and demonstrating the strong defocus applied on M2 to focus the uplink beam and the laser plume. In the figure the slit position at pixel $y = 994$ for spectroscopic observations is shown too.}
              \label{fig:FigRAW}%
    \end{figure}

Figure \ref{fig:FigStrategy} (left) shows the relative positions of the GTC and the WLGSU at the ORM, while Table \ref{TabORM} reports the coordinates for both. The WLGSU and the GTC are about 1113 m away in SW-NE direction, with a difference in elevation (telescope pupil) of about 31 m (Figure \ref{fig:FigStrategy}, right).\\
On the night of the 6th of July the observing strategy was tested, while the actual data whose results we show in this Paper, have been acquired during the night of the 7th of July. The observations have been performed under clear, photometric conditions. The laser was propagated toward the proximity of $\alpha$ Ursae Minoris (Polaris). We have pointed the GTC telescope through the laser plume and the uplink beam at different heights (Figure \ref{fig:FigStrategy}, right) corresponding to 20, 25, 30 and 35 km above sea level. Because of the peculiar observing geometry due to the distance between the GTC and the laser launch telescope, corrected pointing coordinates for the GTC have been calculated using a Python script coded by Fabrice Vidal at Observatoire de Paris Meudon-LESIA.\\
The field rotation was already minimized thanks to the pointing coordinates very close to Polaris, but not enough to perform an observation. For this reason, after every pointing the laser mount and the GTC telescope tracking have been stopped in order to acquire the data and avoid the field rotation due to the observing geometry. We have changed the OSIRIS rotator angle for every acquisition in order to have laser uplink and the plume perpendicular to the slit in Long Slit Spectroscopy, and the same rotator position was used to acquire the Broad Band images too. A strong defocus of the GTC secondary mirror (M2) has been necessary in order to focus the laser plume and the uplink.\\
In Table \ref{tab:ObsStrat} we report a summary of the performed observations. The real-time seeing estimations during the observations reported in the table were retrieved by the Telescopio Nazionale Galileo (TNG) DIMM close to the GTC. In Figure \ref{fig:FigRAW} we show examples of the OSIRIS CCD2 Broad Band raw images of the uplink beam at 20 km (left) and the laser plume (right) with the Sloan R filter. Defocused field stars crossing the OSIRIS FoV can be clearly noticed, revealing the beautiful GTC segmented pupil and demonstrating the strong defocus applied on M2 to focus the uplink beam and the laser plume. The position of the uplink and the plume along of the CCD2 $x$-axis is not random. The targets are placed between pixels $x = 240$ and $x = 350$ which are known to be the most recommended $x$-axis pixels for spectroscopy observations with OSIRIS, minimizing the impact of cosmetics and/or bad pixels along the CCD2 relative columns. In the figure the slit position at pixel $y = 994$ for spectroscopic observations is shown too.\\
Since our data have been acquired after stopping the tracking of the telescope, particular attention has been paid to avoid accidental superimposition on both the laser plume and the uplink beam of field stars crossing the OSIRIS FoV. For that we have imposed some basic rules for what concerns both imaging and spectroscopic data:
\begin{description}
      \item[$-$]  Broad Band imaging
		\begin{enumerate}
			\item[$\bullet$] \textit{Uplink beam}: since the photometric analysis is performed on a $1 \times 1$ arcmin portion of the FoV centered on the uplink beam established altitude, superimposition is allowed in areas outside that portion;
 			\vspace{0.1cm}
			\item[$\bullet$] \textit{Laser plume}: superimposition is never allowed.
		\end{enumerate}
 			\vspace{0.1cm}
     \item[$-$]  Long Slit Spectroscopy: since the position of the slit is along the CCD2 row along pixel $y = 994$ (see Figure \ref{fig:FigRAW}), superimposition is allowed at a distance of $\pm$1 arcmin from the slit for both the uplink beam and the laser plume.
\end{description}
Finally, in order to allow sky subtraction in the data reduction, every science image and spectra have been acquired twice, one with the laser ON, and one with the laser OFF.

\subsection{Data reduction}
Image reduction was performed using a self developed pipeline based on IRAF routines (\citealt{doug93}). Reduction included bias correction, flat fielding and distortion rectification. For the photometric calibration we observed the photometric standard star SA95-193 (\citealt{lan92}).\\
For the spectroscopic data reduction we used again a self developed pipeline based on IRAF routines (\citealt{doug93}). The data reduction included bias and response correction, cosmic ray correction, wavelength calibration, and distortion rectification. The flux calibration was performed using the BD+33-2642 (\citealt{oke90}) spectrophotometric standard star.

\subsection{Photometric calibration}
\begin{figure}
   \centering
   \includegraphics[width=0.49\textwidth]{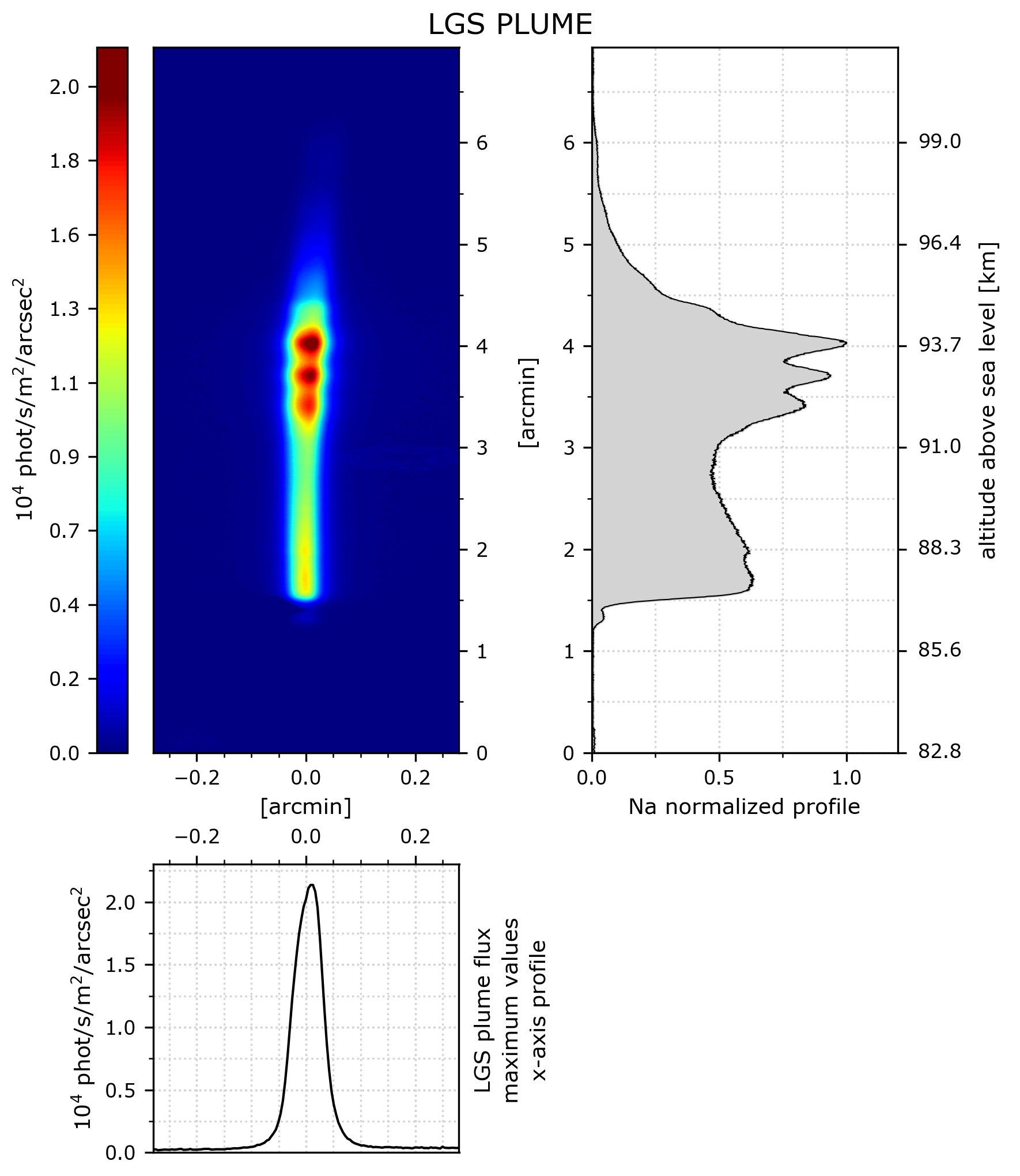}
   \caption{LGS plume form the OSIRIS imager detector for an exposure of 4 sec in R band. The mesospheric sodium normalized profile is shown on the cut on the right, while on the bottom the $x$-axis profile of the flux maximum values is shown.}
              \label{fig:FigLGSprof}%
    \end{figure}

For monochromatic LGS, photometric calibration is not straightforward. We have used the spectrophotometric calibrator BD+33-2642 (\citealt{oke90}) and computed the flux of the LGS D$_{2a}$ line within the 0.6 arcsec slit width, sampled at 0.127 arcsec/px in binning $1 \times 1$. The atmospheric extinction for a wavelength of 589 nm at ORM is 0.09 mag per airmass (\citealt{king85}). The measured OSIRIS magnitude for SA95-193 is m$_{R}$ = 14.9795, while we expect to find m$_{R}$ = 13.9491 at the ground at the star airmass. The above gives an instrument extinction of 1.03 mag in R band, that correspond to an overall throughput of 39\%. From the OSIRIS User Manual\footnote{http://www.gtc.iac.es/instruments/osiris/osiris.php} we know that the detector Quantum Efficiency (QE) is 78\% at (589 $\pm$ 200) nm, while the Sloan R filter transmissivity at 589 nm is 90.4\%. Furthermore, the GTC has an effective collecting area of 73 m$^{2}$. We have converted the SA95-193 total ADU counts into the flux in phot/s/m$^{2}$. We have scaled the fluxes taking into account the difference in spectral behaviour of the system between the Sloan R central wavelength (658 nm) and 589 nm. With this we know how many photons from the LGS are coming from the mesosphere. The location of the slit on the LGS plume, 4.66 arcmin long, was known.\\
From the LGS Broad Band image we have derived the mesospheric sodium profile (Figure \ref{fig:FigLGSprof}, right), of which we know the calibrated flux per pixel at one specific point, as shown on the left of the figure. Integrating the flux scaled along the profile relative intensity, we derived for the observing night a total LGS flux of $1.01 \times 10^{7}$ phot/s/m$^{2}$, which is in accordance with  past photometric measurement campaigns and with the values recorded at the Very Large Telescope Adaptive Optics Facility (VLT AOF) with the Laser Pointing Camera (\citealt{bonac14}; \citealt{cen16}).\\
From these measurements we derived a conversion factor of 0.141 phot/s/m$^{2}$/ADU between the flux above the atmosphere and the OSIRIS detector ADU at 589 nm, at 2.02 airmass (pointing coordinates). Figure \ref{fig:FigLGSprof} (bottom) shows the $x$-axis profile of the flux maximum values. The profile represents the scattering of photons from the LGS plume, and extends up to about $\pm$0.1 arcmin from it.

\section{Results}
\subsection{Raman emission induced by the laser}
\begin{figure*}
         \begin{center}
   $\vcenter{\hbox{\includegraphics[width=0.47\textwidth]{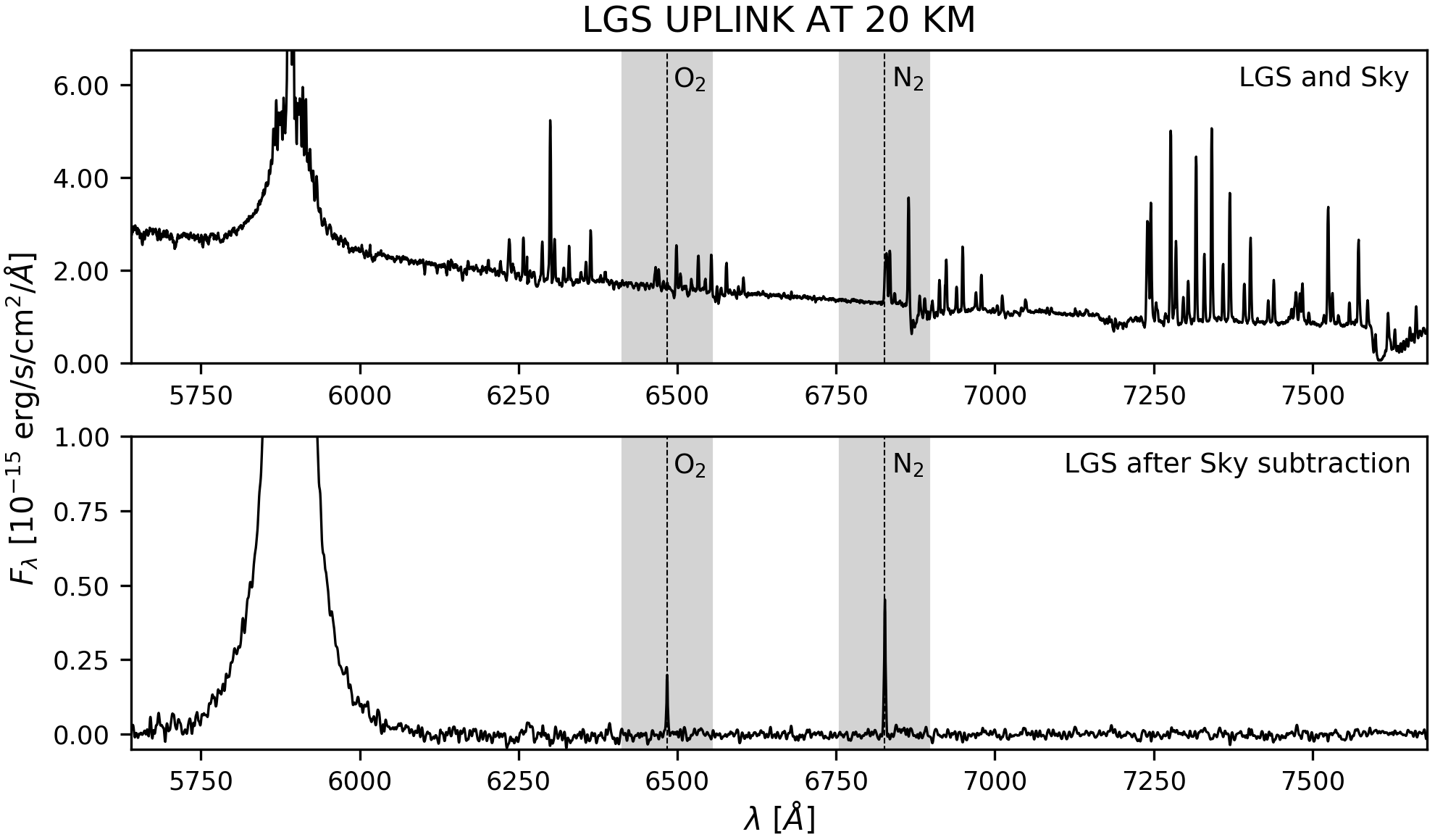}}}$
   \hspace{0.8cm}
   $\vcenter{\hbox{\includegraphics[width=0.47\textwidth]{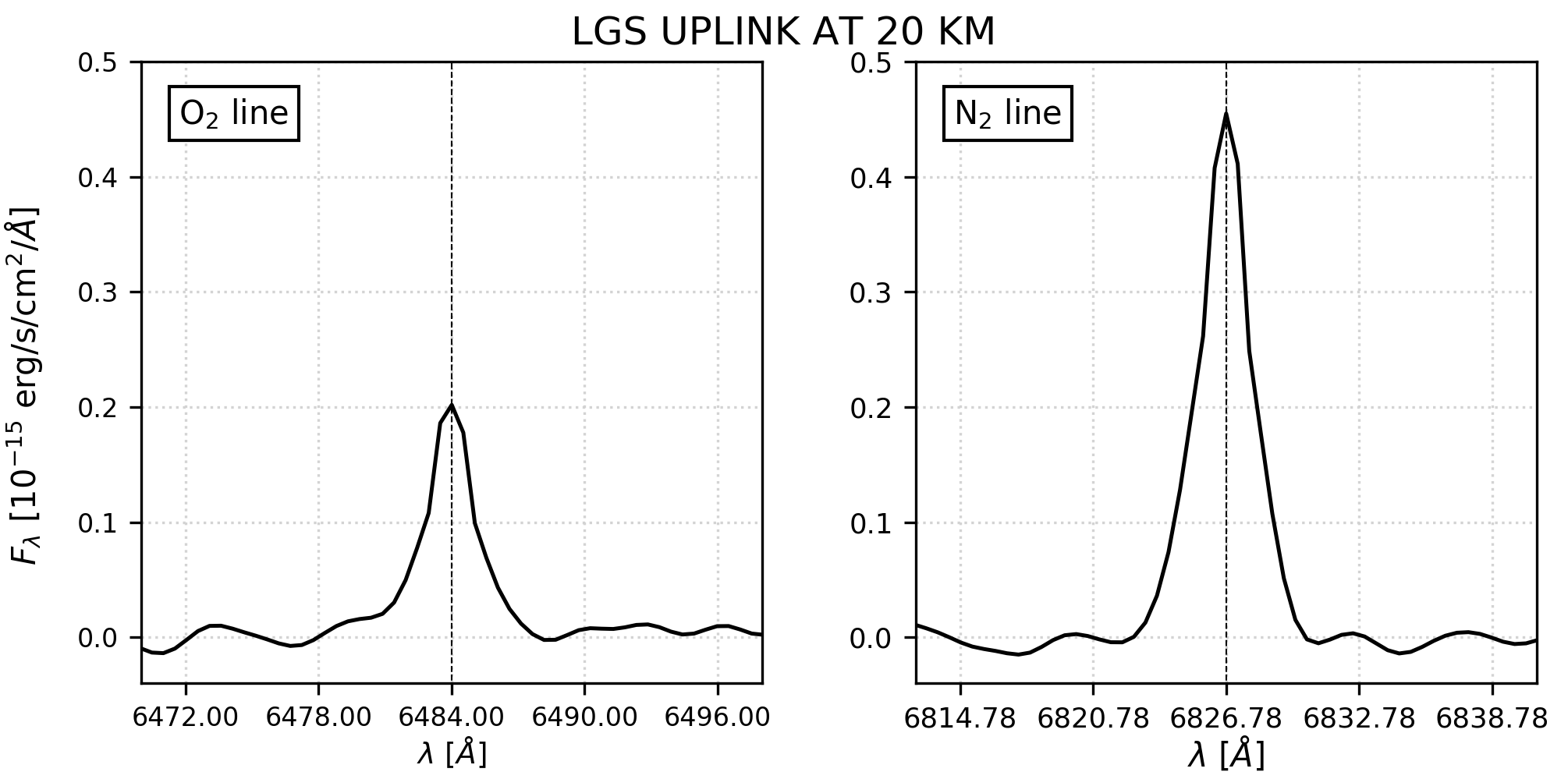}}}$
          \end{center}
   \vspace{-0.4cm}
  \caption{(left) OSIRIS spectra of the uplink beam at 20 km above sea level before (top) and after (bottom) sky subtraction. (right) Detail of the O$_{2}$ and N$_{2}$ Raman emission lines.}
   \label{fig:FigSp20}
 \end{figure*}

\begin{figure*}
         \begin{center}
   $\vcenter{\hbox{\includegraphics[width=0.47\textwidth]{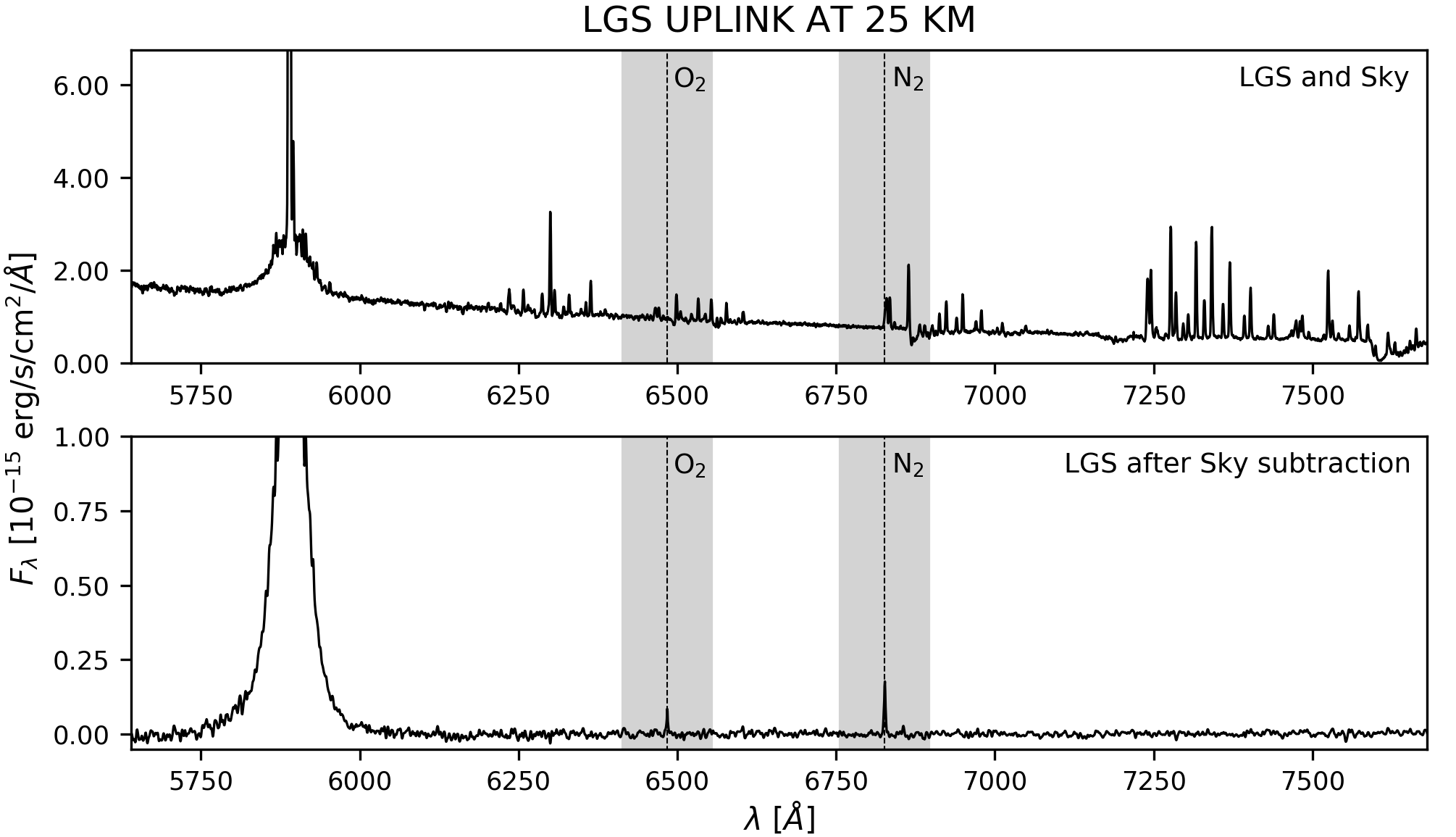}}}$
   \hspace{0.8cm}
   $\vcenter{\hbox{\includegraphics[width=0.47\textwidth]{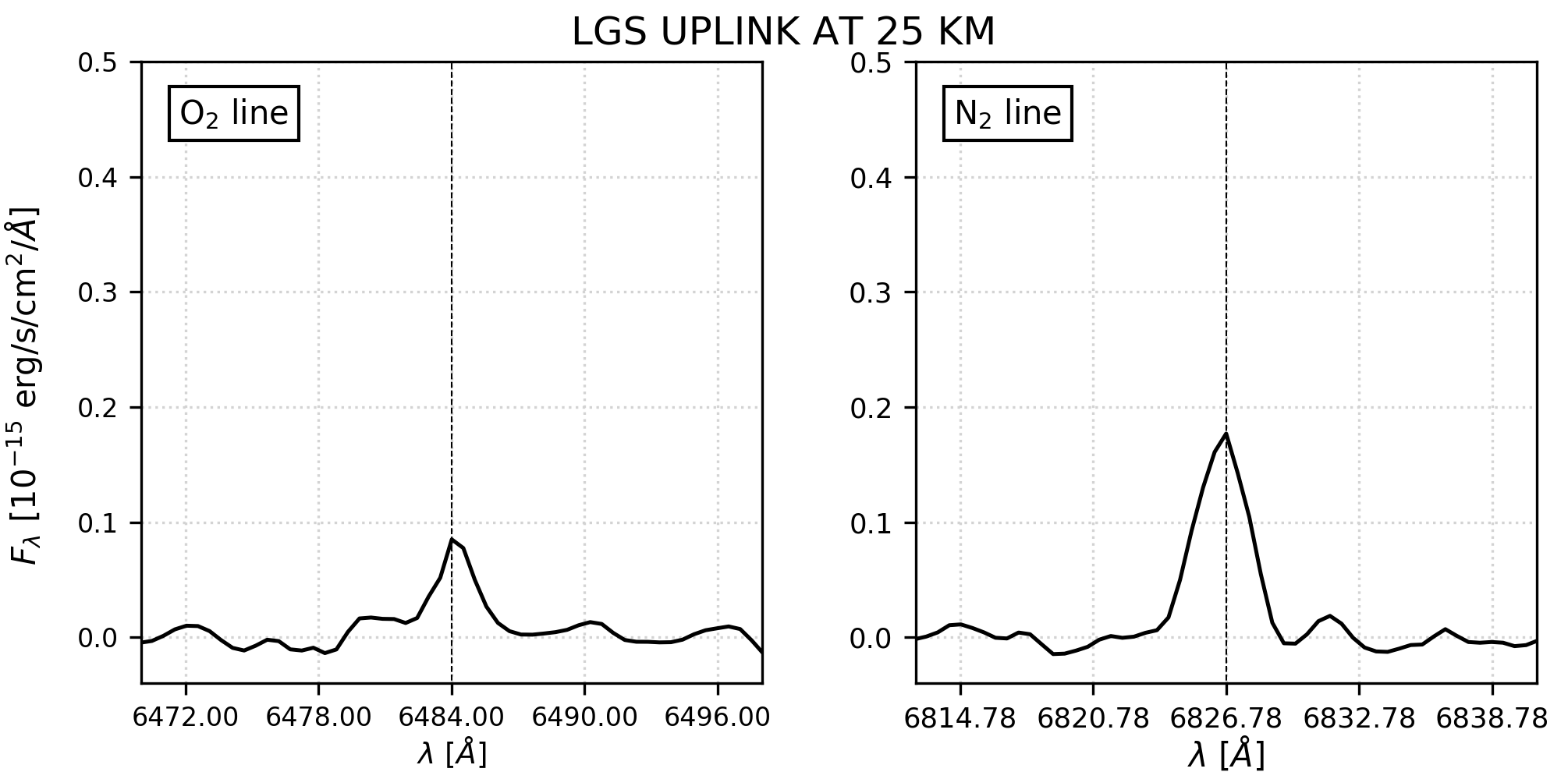}}}$
         \end{center}
   \vspace{-0.4cm}
   \caption{(left) OSIRIS spectra of the uplink beam at 25 km above sea level before (top) and after (bottom) sky subtraction. (right) Detail of the O$_{2}$ and N$_{2}$ Raman emission lines.}
     \label{fig:FigSp25}
 \end{figure*}

\begin{figure*}
         \begin{center}
   $\vcenter{\hbox{\includegraphics[width=0.47\textwidth]{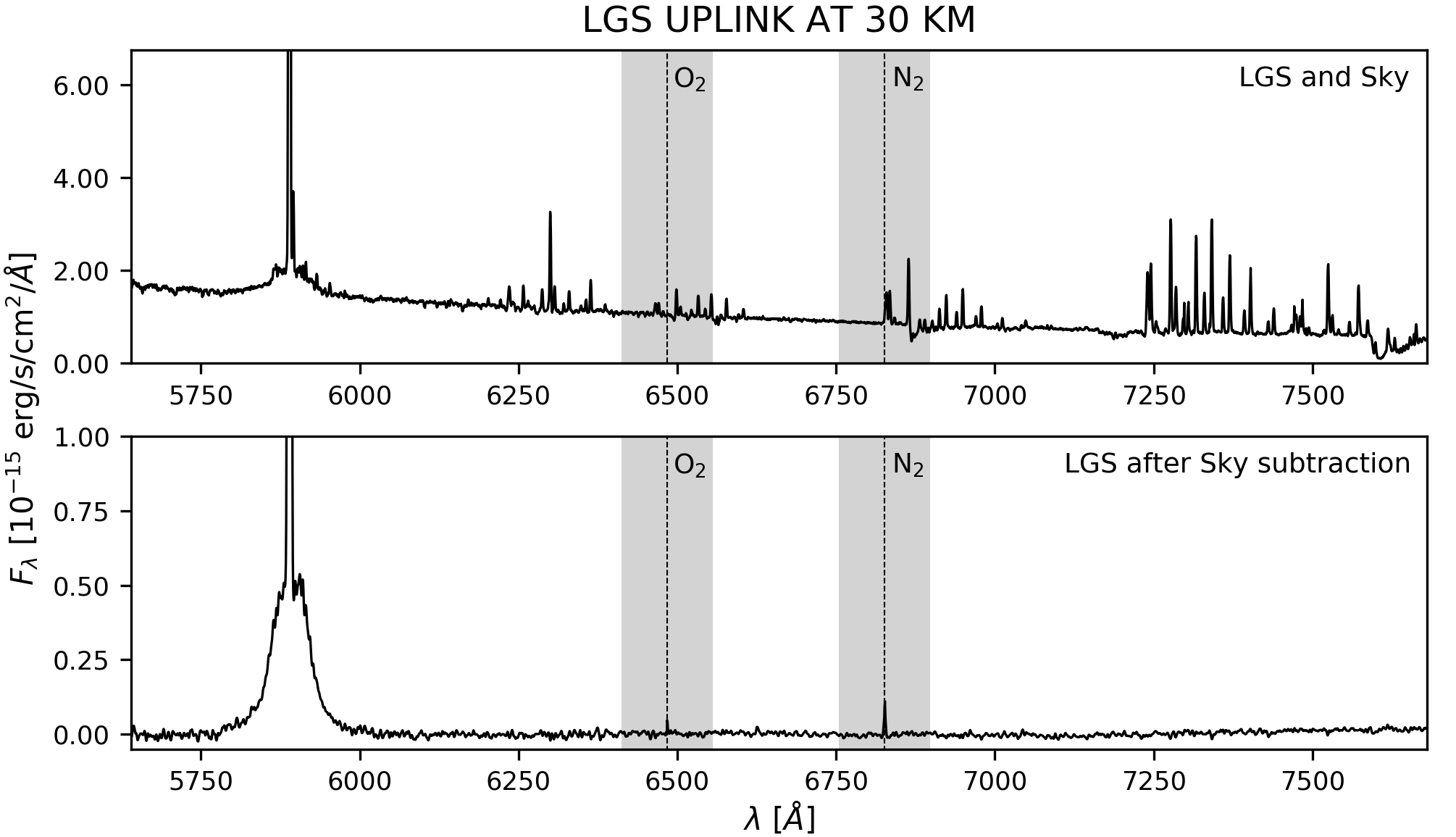}}}$
   \hspace{0.8cm}
   $\vcenter{\hbox{\includegraphics[width=0.47\textwidth]{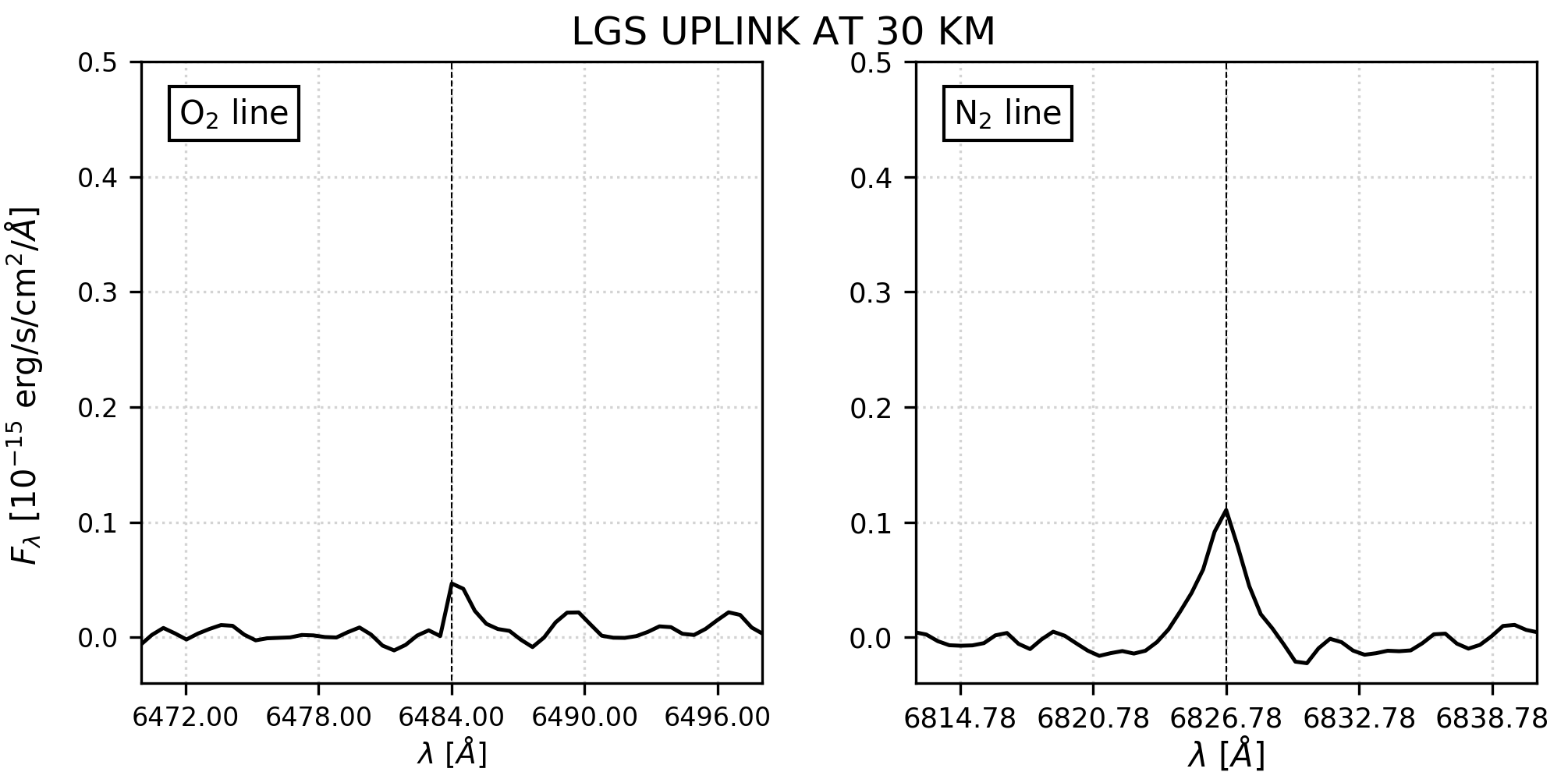}}}$
          \end{center}
    \vspace{-0.4cm}
 \caption{(left) OSIRIS spectra of the uplink beam at 30 km above sea level before (top) and after (bottom) sky subtraction. (right) Detail of the O$_{2}$ and N$_{2}$ Raman emission lines.}
   \label{fig:FigSp30}
 \end{figure*}

\begin{figure*}
         \begin{center}
   $\vcenter{\hbox{\includegraphics[width=0.47\textwidth]{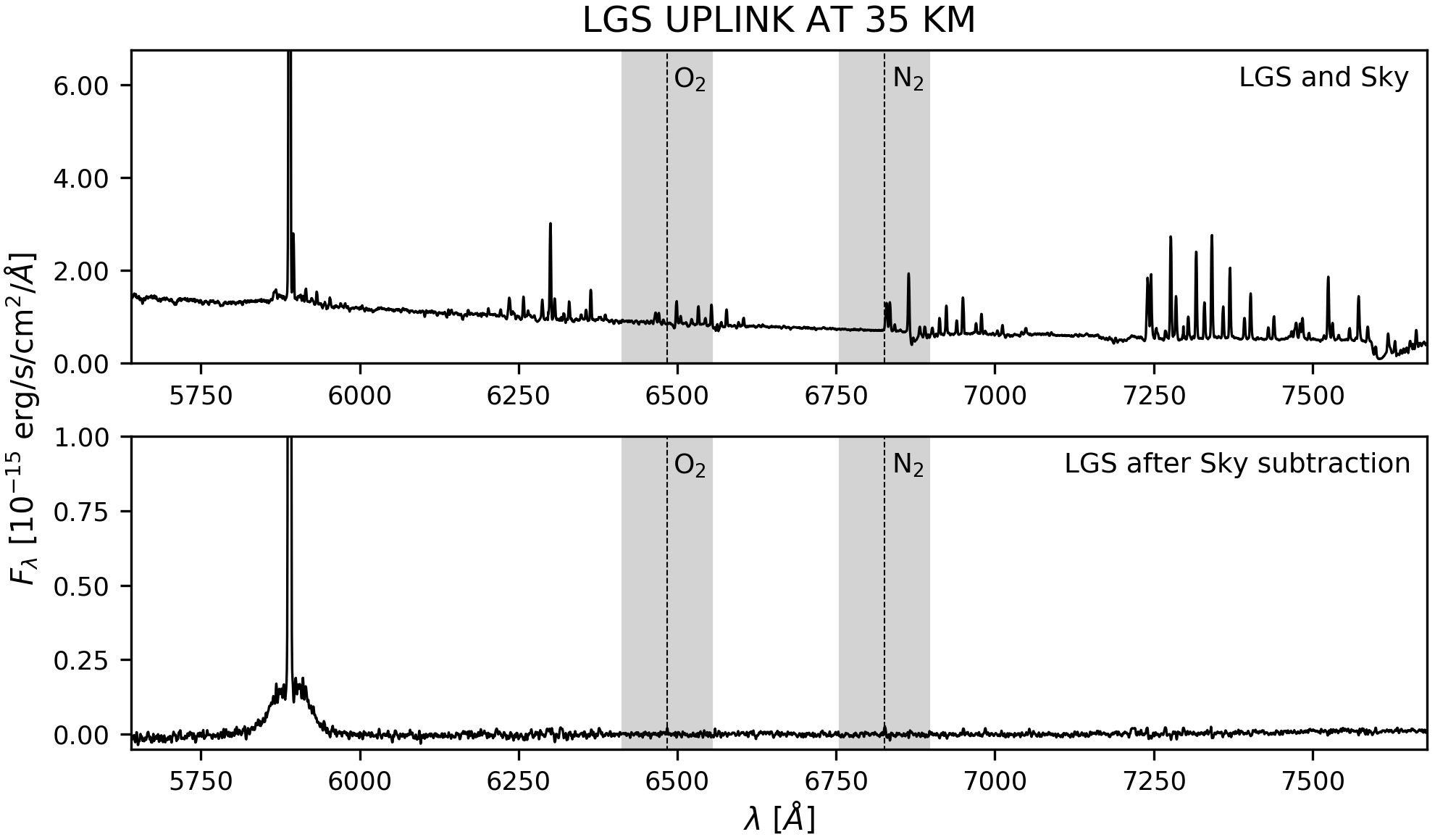}}}$
   \hspace{0.8cm}
   $\vcenter{\hbox{\includegraphics[width=0.47\textwidth]{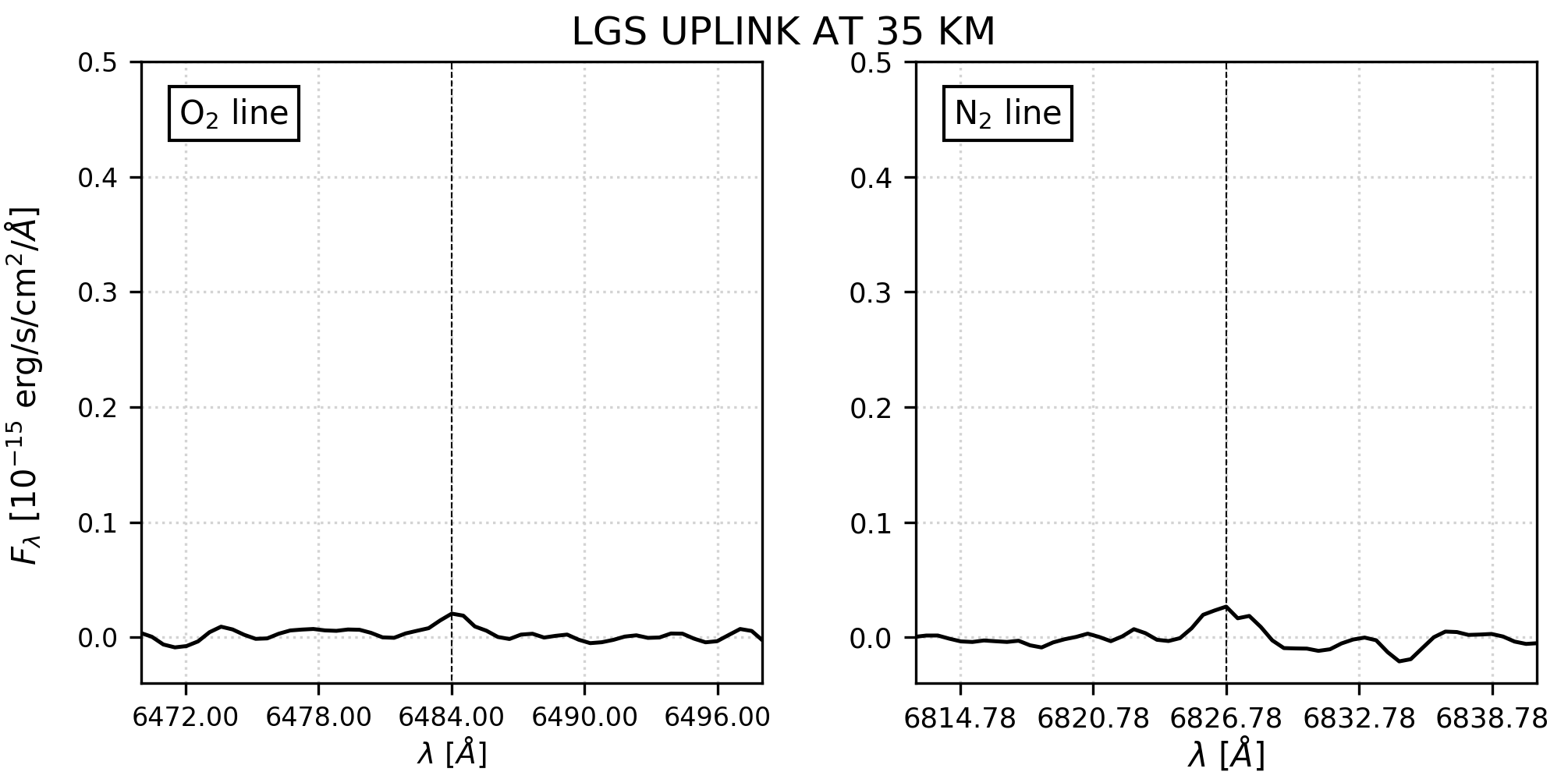}}}$
         \end{center}
   \vspace{-0.4cm}
   \caption{(left) OSIRIS spectra of the uplink beam at 35 km above sea level before (top) and after (bottom) sky subtraction. (right) Detail of the O$_{2}$ and N$_{2}$ Raman emission lines.}
   \label{fig:FigSp35}
 \end{figure*}

Raman scattering is an inelastic scattering of photons by a medium, and this involves vibrational energy being gained by a molecule as incident photons from the laser are shifted to lower energy. When photons are scattered, the majority of them are scattered by Rayleigh scattering (elastic) and have the same energy as the incident photons, but different direction. A smaller fraction of the scattered photons can be scattered inelastically (Raman scattered photons) with a lower energy from the incident ones (\citealt{har78}). Nevertheless, if incident photons collide with an excited molecule (either rotationally or vibrationally), the Raman scattering mechanism can result in an energy gain for the incident photon.\\
Even though our Long Slit Spectroscopy observations have been performed with the highest possible resolution offered by OSIRIS, the setup is not sufficient to reveal the entire rotational-vibrational forests of molecular lines of the Raman emission. What we could observe are the unresolved Q-branch of the O$_{2}$ and N$_{2}$ molecules first vibrational transition. Having acquired longer exposures or used instruments with higher resolution would have revealed affection by R- and S-branch too, together with lines transitions associated to other molecules like H$_{2}$O, CH$_{4}$, CO$_{2}$, etc. (for example see \citealt{vogt19}). However, for simplicity from now on, we will refer to our detected unresolved Q-branch of the first vibrational transition from the O$_{2}$ and N$_{2}$ molecules simply as the "O$_{2}$ line" and the "N$_{2}$ line".\\
Figures \ref{fig:FigSp20}, \ref{fig:FigSp25}, \ref{fig:FigSp30}, \ref{fig:FigSp35} show on the right the obtained spectra at the respective altitudes before and after sky subtraction. On the left of every figure the detail of the O$_{2}$ and N$_{2}$ Raman emissions are shown. In the figures, the pure rotational Raman lines located in the immediate vicinity of the LGS line are clearly visible. The integrated fluxes $F_{\lambda}$ for the LGS, and the O$_{2}$ and N$_{2}$ Raman emission lines are reported in Table \ref{TabIntFlux}. In the table, the ratios of the measured LGS flux with respect to respective O$_{2}$ and N$_{2}$ lines fluxes are reported too. The calculated $F_{\lambda}$ are comparable to those found at 25 km at Paranal Observatory by \citet{vogt17}.\\
We observed the laser uplink at 20, 25, 30 and 35 km above sea level. As expected, at 20 km we detected the strongest emission. The emission progressively weakens with the altitude until it becomes undetectable above 30 km. The Raman emission lines are comparable in intensity to other telluric lines thus they should be removed during standard spectra data reduction, nevertheless photon noise remnant is not quantified at this stage.\\
For low-resolution spectrographs like OSIRIS, the 1.713 GHz separated  D$_{2a}$ and  D$_{2b}$ laser emission lines (1.6 MHz FWHM) forming the LGS, can be approximated to a Dirac Delta, thus the spectrum observed is indeed the spectrograph spectral transfer function at 589 nm. This means that the LGS line may be used online as spectral transfer function calibrator. Finally, knowing the Raman lines, together with the laser main line, they can be used for the wavelength calibration of the spectrograph.
\begin{figure*}
         \begin{center}
   $\vcenter{\hbox{\includegraphics[width=0.33\textwidth]{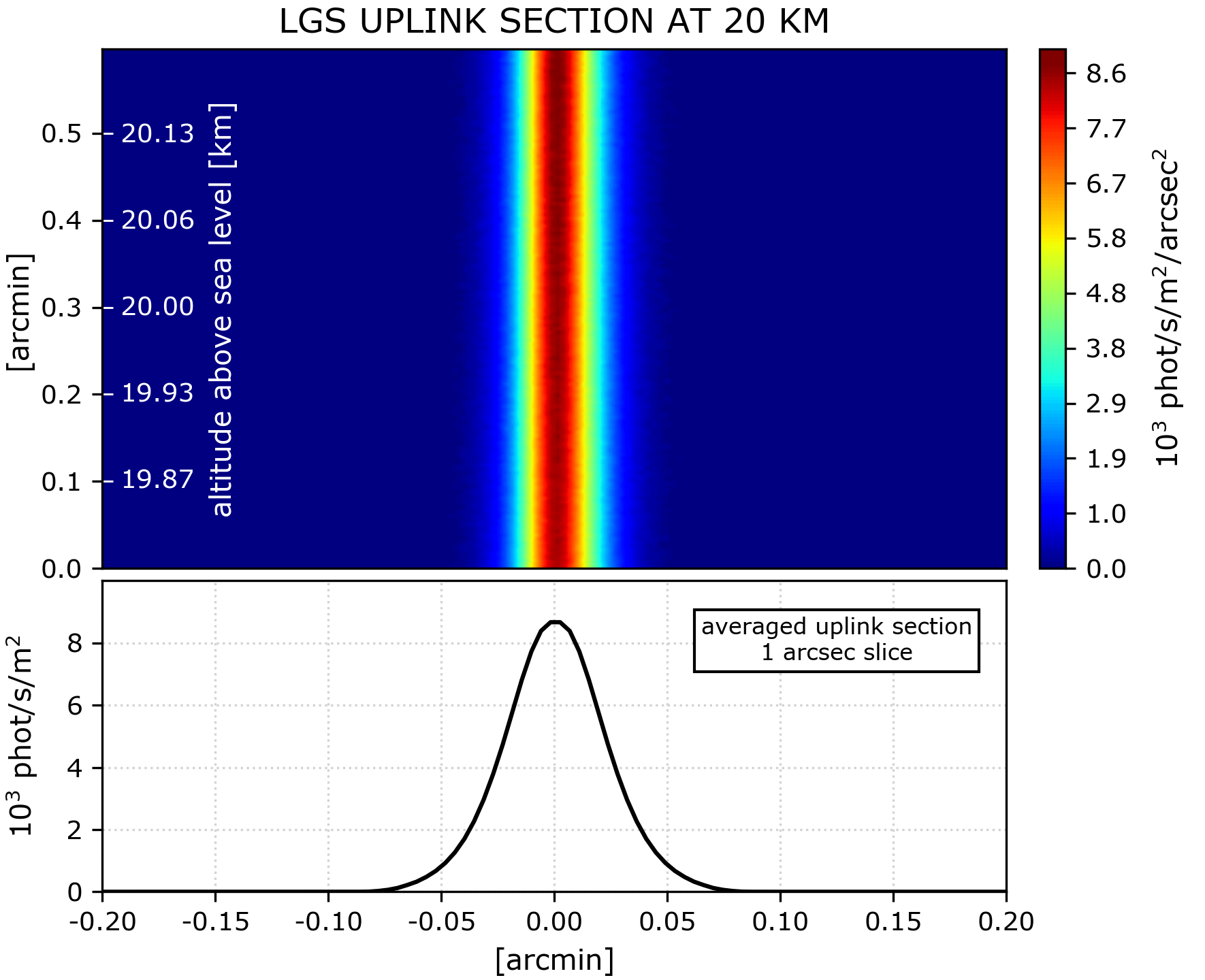}}}$
   \hspace{1.8cm}
   $\vcenter{\hbox{\includegraphics[width=0.33\textwidth]{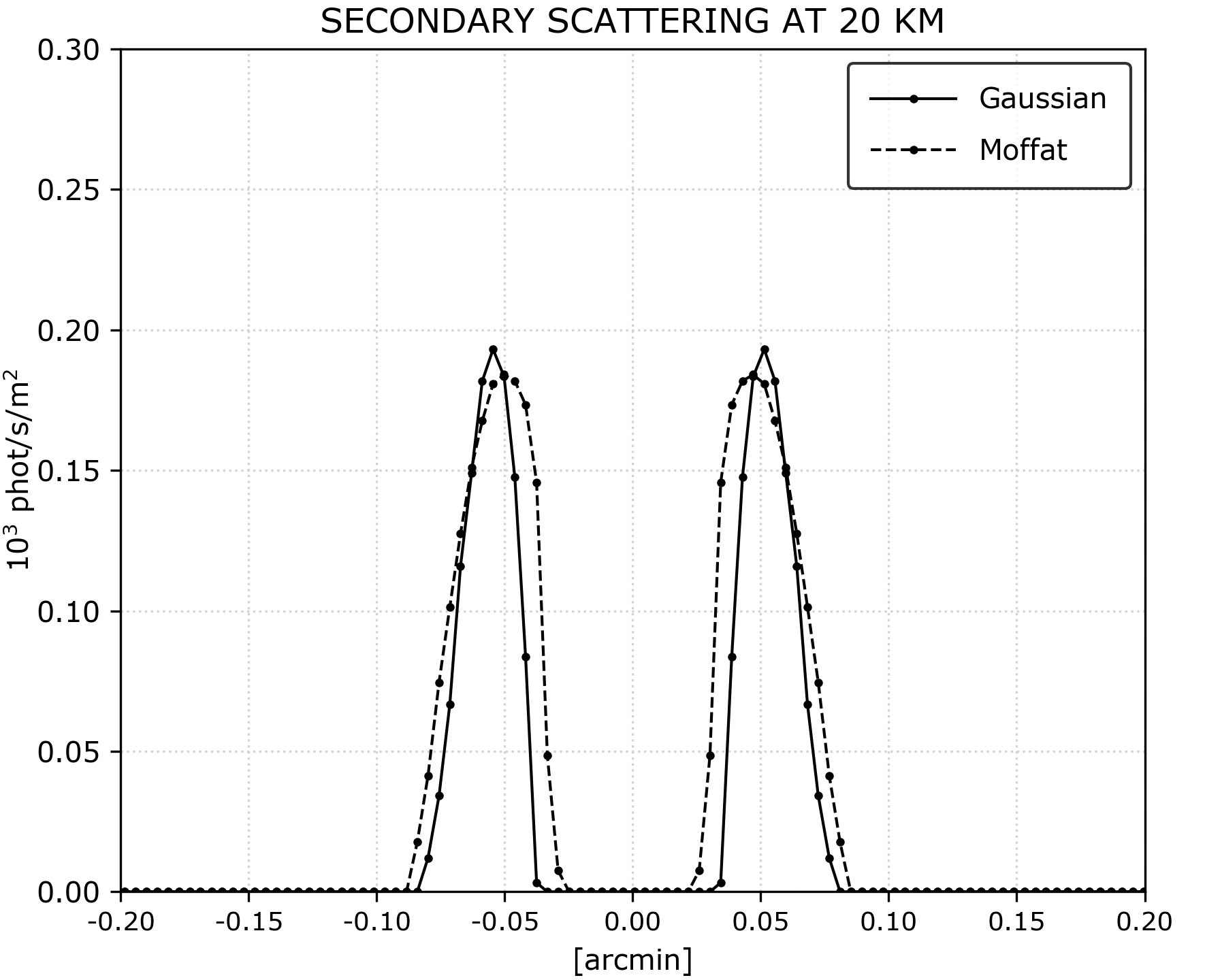}}}$
          \end{center}
   \vspace{-0.45cm}
  \caption{(left) Uplink section at 20 km above sea level and the averaged uplink section from 1 arcsec slice. (right) Secondary scattering from the difference between the Moffat or Gaussian distributions with respect to the averaged uplink section.}
   \label{fig:FigLGS20}
 \end{figure*}
\begin{figure*}
         \begin{center}
   $\vcenter{\hbox{\includegraphics[width=0.33\textwidth]{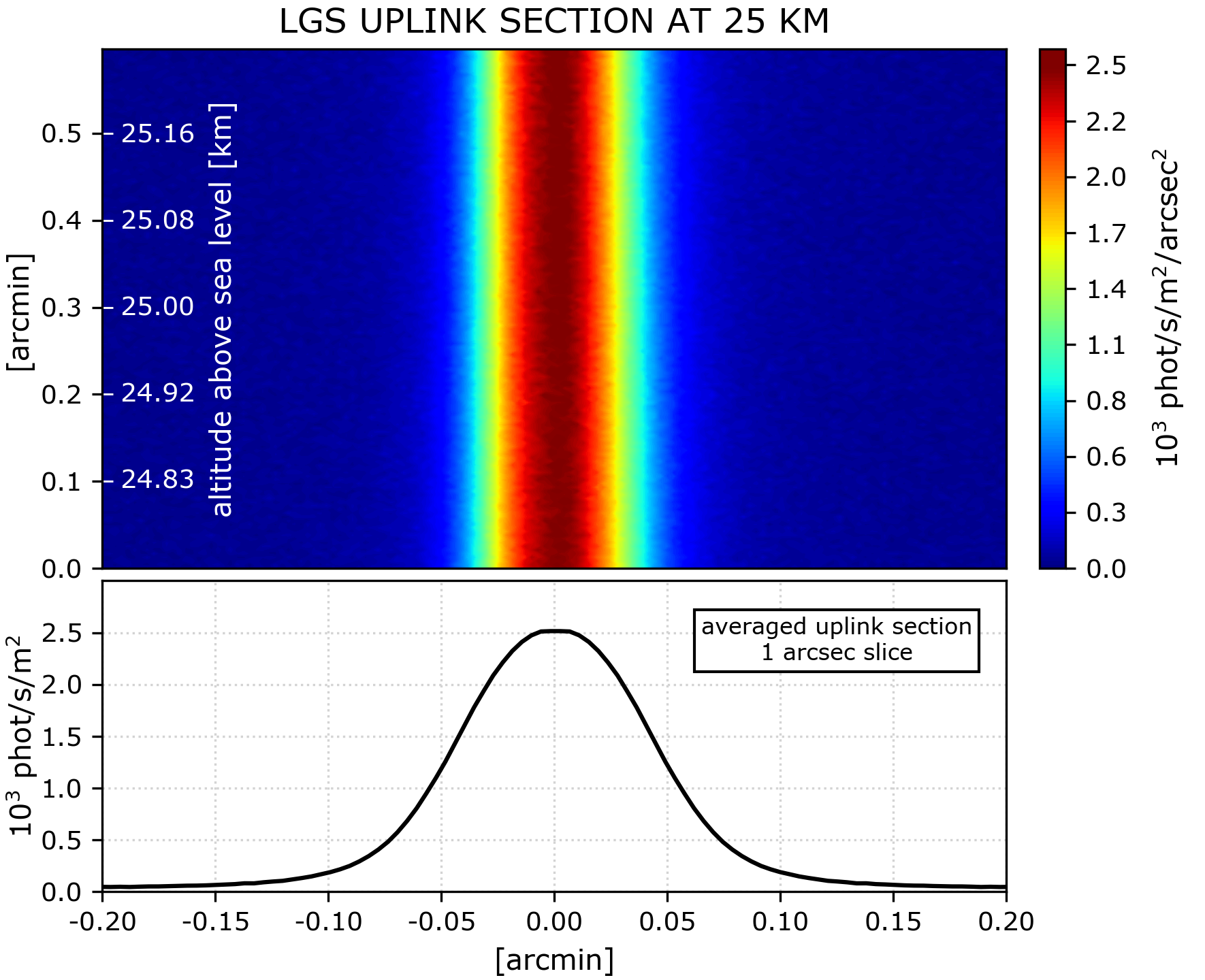}}}$
   \hspace{1.8cm}
   $\vcenter{\hbox{\includegraphics[width=0.33\textwidth]{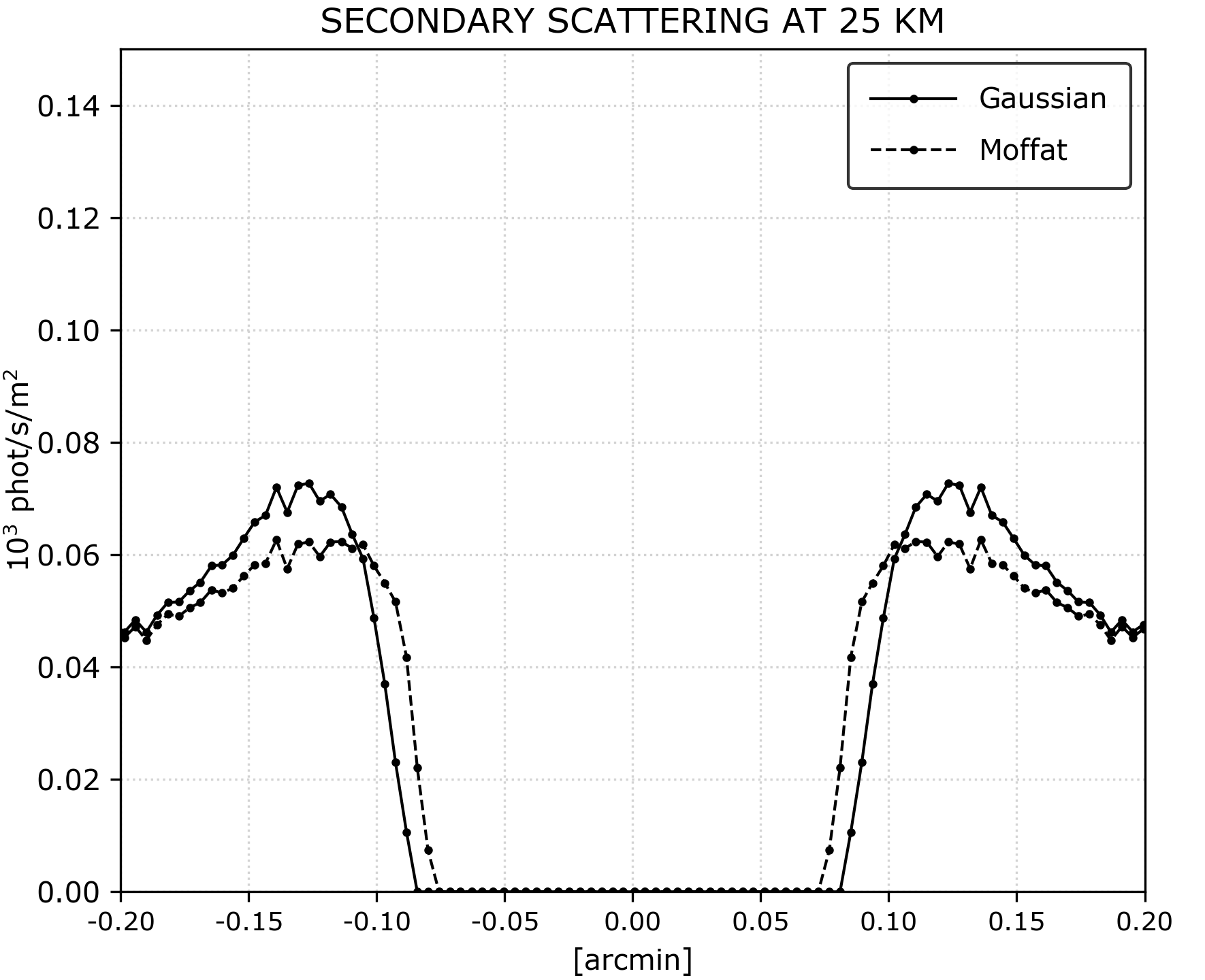}}}$
         \end{center}
    \vspace{-0.45cm}
 \caption{(left) Uplink section at 25 km above sea level and the averaged uplink section from 1 arcsec slice. (right) Secondary scattering from the difference between the Moffat or Gaussian distributions with respect to the averaged uplink section.}
     \label{fig:FigLGS25}
 \end{figure*}
\begin{figure*}
         \begin{center}
   $\vcenter{\hbox{\includegraphics[width=0.33\textwidth]{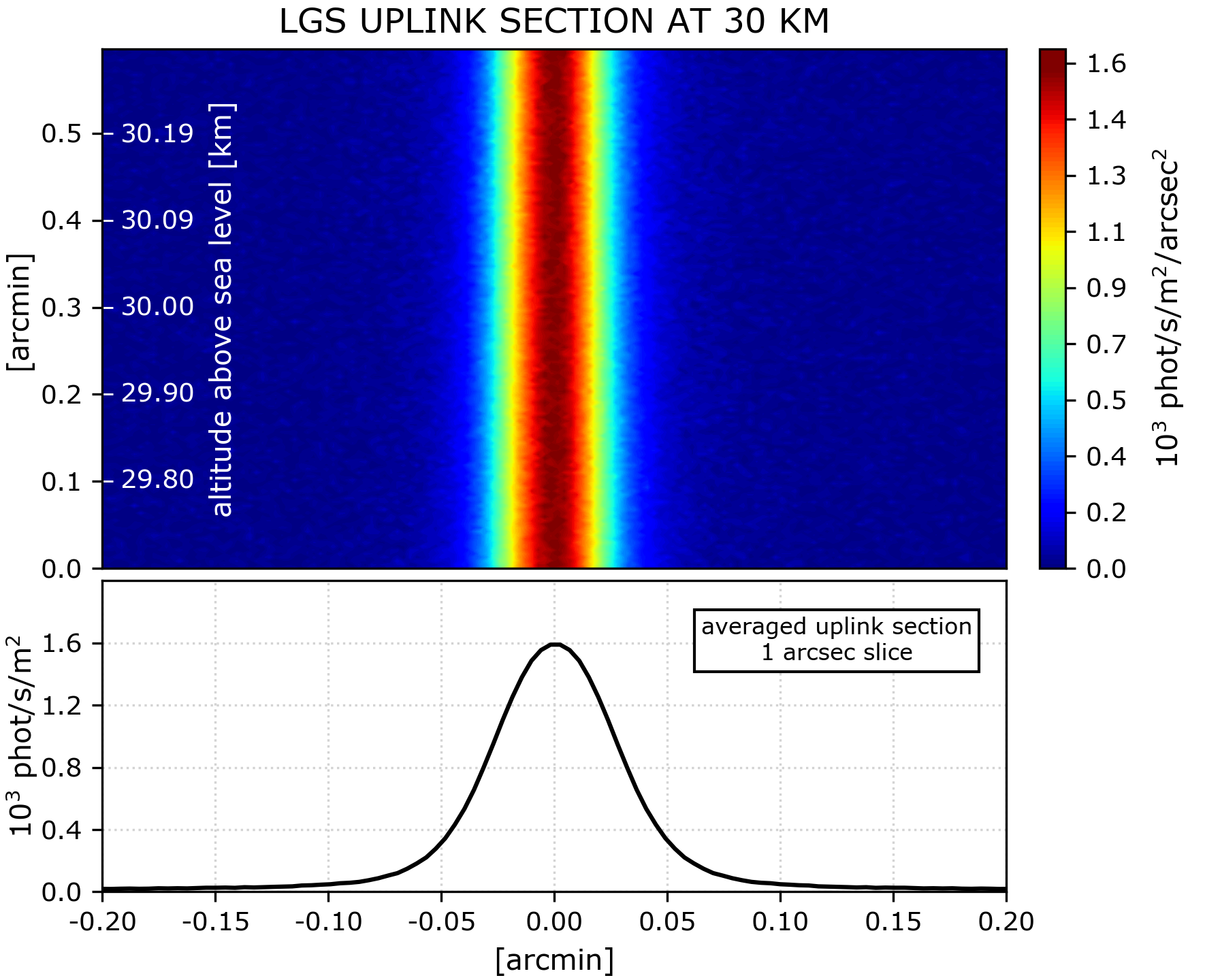}}}$
   \hspace{1.8cm}
   $\vcenter{\hbox{\includegraphics[width=0.33\textwidth]{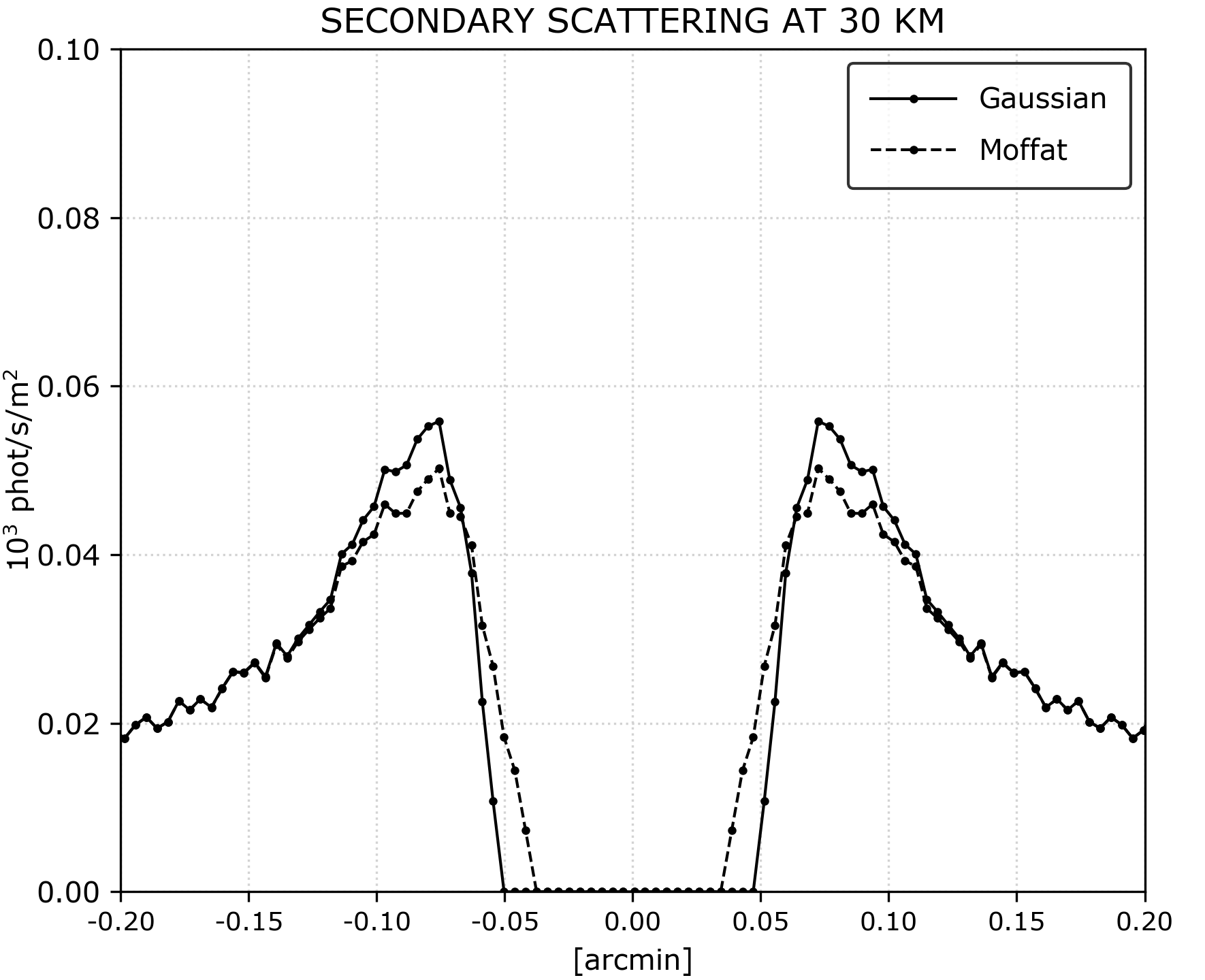}}}$
          \end{center}
   \vspace{-0.45cm}
  \caption{(left) Uplink section at 30 km above sea level and the averaged uplink section from 1 arcsec slice. (right) Secondary scattering from the difference between the Moffat or Gaussian distributions with respect to the averaged uplink section.}
   \label{fig:FigLGS30}
 \end{figure*}
\begin{figure*}
         \begin{center}
   $\vcenter{\hbox{\includegraphics[width=0.33\textwidth]{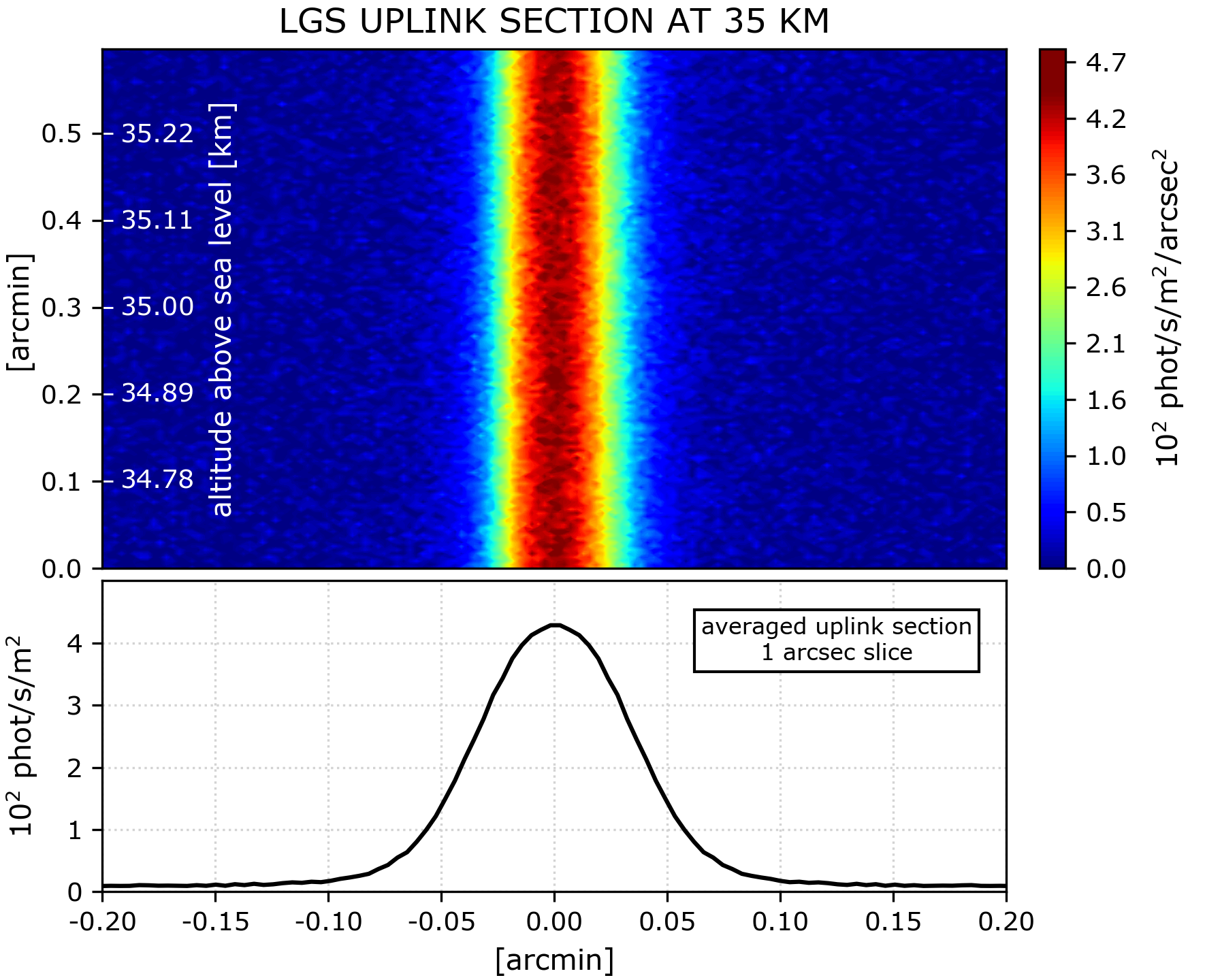}}}$
   \hspace{1.8cm}
   $\vcenter{\hbox{\includegraphics[width=0.33\textwidth]{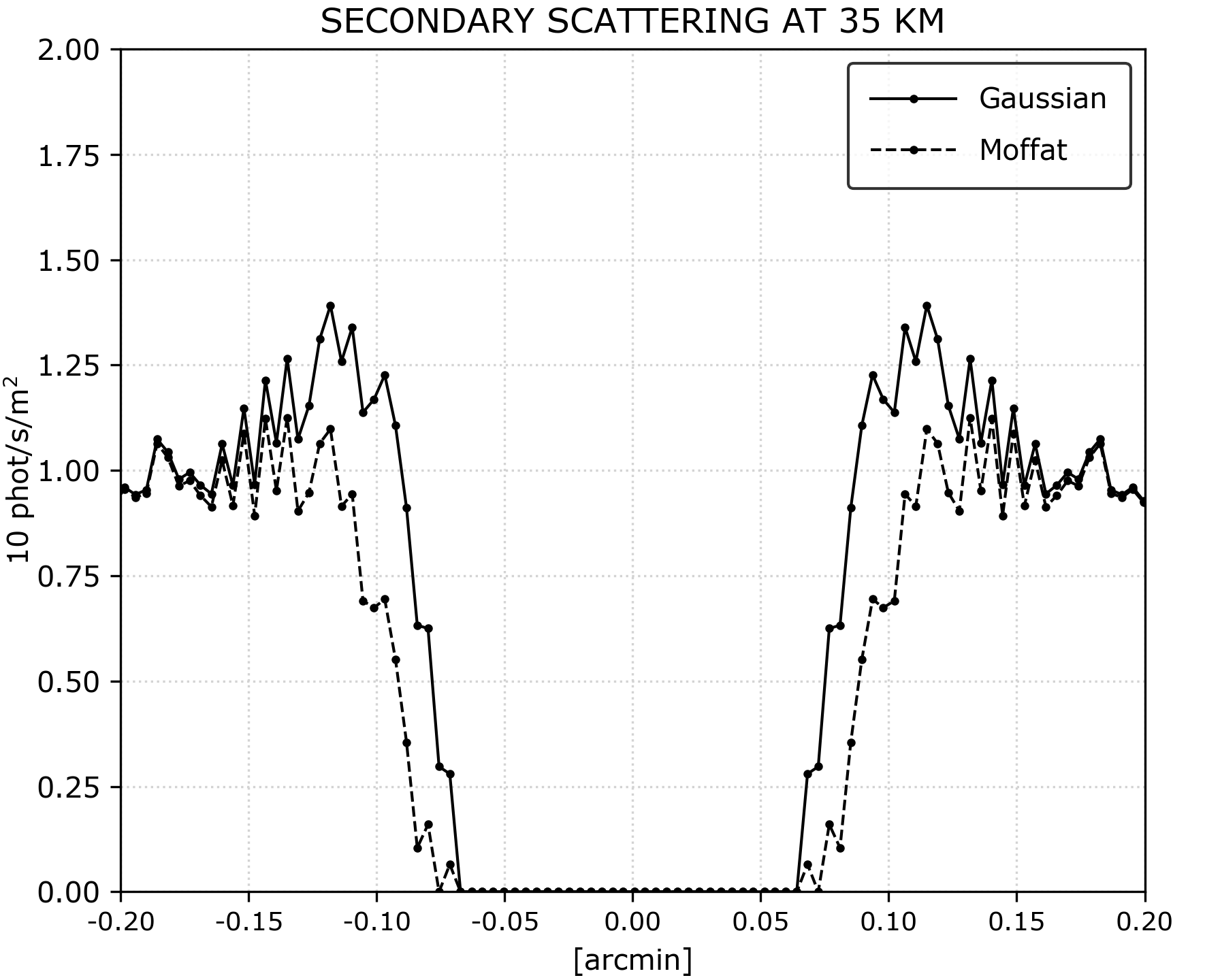}}}$
         \end{center}
   \vspace{-0.45cm}
  \caption{(left) Uplink section at 35 km above sea level and the averaged uplink section from 1 arcsec slice. (right) Secondary scattering from the difference between the Moffat or Gaussian distributions with respect to the averaged uplink section.}
   \label{fig:FigLGS35}
 \end{figure*}

\begin{table*}
\small
      \caption[]{$F_{\lambda}$ in 10$^{-20}$ erg/s/cm$^{2}$/W for the LGS, and the O$_{2}$ and N$_{2}$ Raman emission lines. The ratio of the measured LGS flux with respect to respective O$_{2}$ and N$_{2}$ fluxes is reported too. }
         \label{TabIntFlux}
         \begin{center}
         \begin{tabular}{ccccccccc}
            \hline
              \textbf{Altitude} &   &  \textbf{LGS} & & \textbf{O$_{2}$} & \textbf{Ratio} & & \textbf{N$_{2}$} & \textbf{Ratio}\\
                  & & 5889.96 \r{A} & & 6484.00 \r{A} & \textbf{O$_{2}$/LGS} & & 6826.78 \r{A} & \textbf{N$_{2}$/LGS}\\
            \hline
            20 km & & $3.31 \times 10^{7}$ & & $1.23 \times 10^{4}$ & $\sim$$3.7 \times 10^{-4}$ & & $2.37 \times 10^{4}$ & $\sim$$7.2 \times 10^{-4}$\\
            \noalign{\smallskip}
            25 km & & $1.40 \times 10^{7}$ & & $7.55 \times 10^{3}$ & $\sim$$5.4 \times 10^{-4}$ & & $1.20 \times 10^{4}$ & $\sim$$8.6 \times 10^{-4}$\\
            \noalign{\smallskip}
            30 km & & $6.18 \times 10^{6}$ & & $6.22 \times 10^{3}$ &  $\sim$$1.0 \times 10^{-3}$ & & $6.68 \times 10^{3}$ & $\sim$$1.1 \times 10^{-3}$\\
            \noalign{\smallskip}
            35 km & & $2.10 \times 10^{6}$ & & $2.68 \times 10^{3}$ & $\sim$$1.3 \times 10^{-3}$ & & $3.26 \times 10^{3}$ & $\sim$$1.6 \times 10^{-2}$\\
            \hline
         \end{tabular}
         \end{center}
   \end{table*}
\begin{table}
\small
      \caption[]{Flux in phot/s/m$^{2}$/arcsec$^{2}$ and phot/s/ster of the laser uplink beam at different altitudes and the respective distance from the source.}
         \label{TabSterFlux}
         \begin{center}
         \begin{tabular}{c c c c}
            \hline
              \textbf{Altitude} & \textbf{Distance} & \textbf{Flux} & \textbf{Radiance}\\
               km   & km  & phot/s/m$^{2}$/arcsec$^{2}$ & phot/s/ster \\
            \hline
            20 &  41.88 & $8.67 \times 10^{3}$ & $1.52 \times 10^{13}$ \\
            \noalign{\smallskip}
            25 &  51.80 & $2.52 \times 10^{3}$ & $0.68 \times 10^{13}$ \\
            \noalign{\smallskip}
            30 &  61.57 & $1.59 \times 10^{3}$ & $0.60 \times 10^{13}$ \\
            \noalign{\smallskip}
            35 &  71.18 & $0.43 \times 10^{3}$ & $0.22 \times 10^{13}$ \\
            \hline
         \end{tabular}
         \end{center}
   \end{table}

\subsection{Photometry of the laser beam scattering}\label{laserphot}
On the left of Figures \ref{fig:FigLGS20}, \ref{fig:FigLGS25}, \ref{fig:FigLGS30}, \ref{fig:FigLGS35} we show the laser uplink sections at 20, 25, 30 and 35 km above sea level together with the averaged uplink section from a 1 arcsec slice. We clearly notice a difference in the FWHM of slices at different altitudes. This difference is ascribed to variation in scattering emission with altitude as well as a non perfect focusing of the GTC M2 on the uplink and to a non negligible variable seeing during the night of observation (for example this is recognizable in the wider FWHM of the uplink section at 25 km in Figure \ref{fig:FigLGS25} when we experienced a spike in the seeing, see Table \ref{tab:ObsStrat}). The imperfect focusing and the variable seeing are not a real issue because in real life scenarios telescopes observing and eventually colliding with the laser uplink will not be in focus on the uplink. The wider scattering of the photons around the uplink horizontal section peak will make the area affected by the uplink larger or thinner. The figures puts in evidence that for our setup and with the telescope in focus on the uplink, crossing the laser uplink at a distance lower than $\pm$0.2 arcmin will produce a contamination of the image. A telescope focused at infinity during real astronomical observations will already be affected at larger distances from the beam because of the wider profile of a defocused uplink section, albeit with less photon angular density.\\
\begin{figure}
         \begin{center}
	\includegraphics[width=0.47\textwidth]{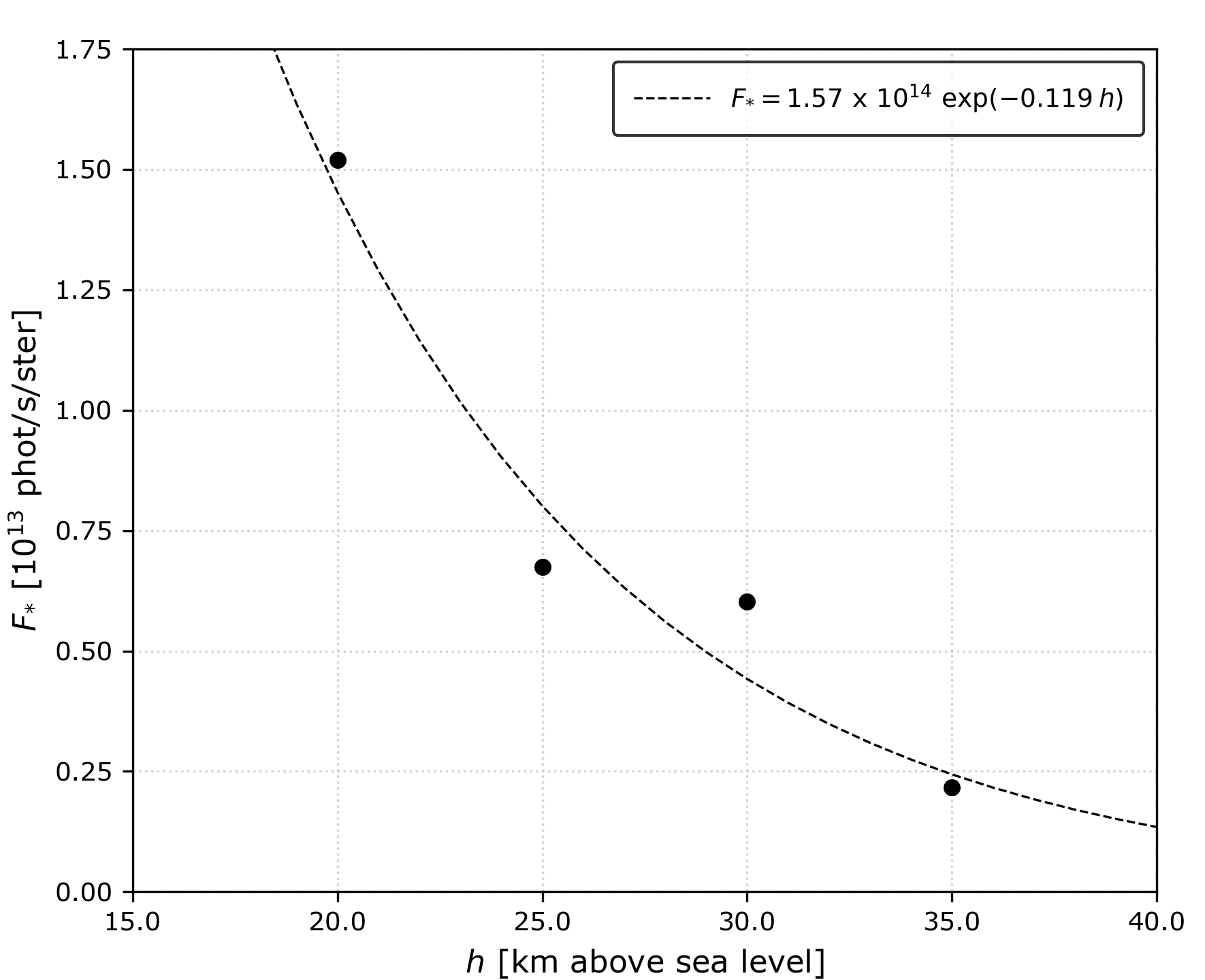}
         \end{center}
   \vspace{-0.45cm}
  \caption{Trendline of the uplink beam scattering radiance (flux per second per steradian).}
   \label{fig:FigSter}
 \end{figure}
The case is different for what concerns the LGS itself. The laser produces an artificial star of magnitude around 7 in the visible. From Figure \ref{fig:FigLGSprof} (bottom) we confirm that the scattering extends up to about $\pm$0.1 arcmin from the plume, diminishing in this way the safety angular distance to avoid contamination of the science data from the 589 nm photons. This may be considered a more solid statement, because the difference in focus from the artificial star produced in the mesosphere and the infinity will be less significant, although depending on the telescope size used. This keeps more significance when considering that our experimental setup simulates telescopes far from the origin of the laser propagation, intercepting the LGS at a linear distance significantly higher than 90 km.\\
In presence of primary scattering only, i.e. scattering generated directly by the laser uplink photons, the shape of the horizontal section of the uplink beam must be Gaussian. In case of presence of secondary scattering (generated by primary scattering photons, outside and in the vicinity of the uplink laser beam) the wings of the distribution will widen loosing the Gaussian distribution. Indeed, we have verified that the shapes of the uplink beam section at different altitudes reported on the left of Figures \ref{fig:FigLGS20}, \ref{fig:FigLGS25}, \ref{fig:FigLGS30}, \ref{fig:FigLGS35} are not Gaussian, thus their difference with a Gaussian or Moffat distribution will put in evidence the secondary scattering as shown on the right of the figures.\\
Integrating the area below the uplink beam section we get the solid angle flux of the uplink for a specific altitude in phot/s/m$^{2}$/arcsec$^{2}$.\\
The solid angle flux deriving from scattering of the uplink beam is useful as normalized parameter, independent on the location and aperture of a receiver telescope, allowing the light contamination calculations applicable to every telescope facility in the same observatory at a given time. Indeed a calibration for different aerosol and dust conditions obtained in long term data gathering at LGS-AO facilities would allow to establish a model and define the minimum distance between a telescope line-of-sight and the uplink laser beams, a key parameter to be inserted in the automatic Laser Traffic Control System software. Table \ref{TabSterFlux} reports the calculated fluxes in phot/s/m$^{2}$/arcsec$^{2}$ and the radiance in phot/s/ster, while in Figure \ref{fig:FigSter} the radiance are plotted against the respective altitude above sea level. The exponential fit of the trend of the radiance in altitude is:
\begin{equation}
F_{*} = 1.57 \times 10^{14} \exp(-0.119 \: h)
\end{equation}
where $F_{*}$ is the radiance in phot/s/ster an $h$ the relative altitude above sea level in km.\\
Our results are based on observations performed during photometric nights free of dust. We do expect an increase of the primary and secondary scattering levels in presence of aerosols, but those effects could not be observed and quantified yet (see Section \ref{aerosol}).\\
\begin{table*}
      \caption[]{Redshift ranges at which important spectral lines coincide  with the LGS, the O$_{2}$ and N$_{2}$ vibrational lines and their respective rotational side-lines.}
         \label{TabRedshifts}
        \begin{center}
         \begin{tabular}{c c c c r c r c r}
            \hline
            \noalign{\smallskip}
                          & &                   & & \multicolumn{5}{c}{\textbf{Redshift range}}\\
            \noalign{\smallskip}
            \cline{5-9}
            \noalign{\smallskip}
            \textbf{Line} & & \textbf{Restframe} & & \multicolumn{1}{c}{\textbf{LGS}} &  & \multicolumn{1}{c}{\textbf{O$_{2}$}} & & \multicolumn{1}{c}{\textbf{N$_{2}$}}\\
                          &  & $\lambda$ [\r{A}] & & \multicolumn{1}{c}{5889.96 \r{A}} & & \multicolumn{1}{c}{6484.00 \r{A}}  & & \multicolumn{1}{c}{6826.78 \r{A}} \\
            \hline
            \noalign{\smallskip}
            \texttt{Ly$\alpha$}  & & \textbf{1215.67}  & & 3.845 $\pm$ 0.082 & & 4.334 $\pm$ 0.041 & & 4.616 $\pm$ 0.062 \\
            \noalign{\smallskip}
            \texttt{[OII]}     & & \textbf{3727.3}   & & 0.580 $\pm$ 0.027 & & 0.740 $\pm$ 0.013 & & 0.832 $\pm$ 0.020 \\
            \noalign{\smallskip}
            \texttt{Ca(H)}     & & \textbf{3933.7}   & & 0.497 $\pm$ 0.025 & & 0.648 $\pm$ 0.013 & & 0.735 $\pm$ 0.019 \\
            \noalign{\smallskip}
            \texttt{Ca(K)}     & & \textbf{3968.5}   & & 0.484 $\pm$ 0.025 & & 0.634 $\pm$ 0.013 & & 0.720 $\pm$ 0.019 \\
            \noalign{\smallskip}
            \texttt{H$\delta$}   & & \textbf{4102.8}   & & 0.436 $\pm$ 0.024 & & 0.580 $\pm$ 0.012 & & 0.664 $\pm$ 0.018 \\
            \noalign{\smallskip}
            \texttt{G-band} & & \textbf{4304.4}   & & 0.368 $\pm$ 0.023 & & 0.506 $\pm$ 0.012 & & 0.586 $\pm$ 0.017 \\
            \noalign{\smallskip}
            \texttt{H$\gamma$}   & & \textbf{4340.0}   & & 0.357 $\pm$ 0.023 & & 0.494 $\pm$ 0.012 & & 0.573 $\pm$ 0.017 \\
            \noalign{\smallskip}
            \texttt{H$\beta$}    & & \textbf{4861.3}   & & 0.212 $\pm$ 0.021 & & 0.334 $\pm$ 0.010 & & 0.404 $\pm$ 0.015 \\
            \noalign{\smallskip}
            \texttt{[OIII]}    & & \textbf{4959.0}   & & 0.188 $\pm$ 0.020 & & 0.308 $\pm$ 0.010 & & 0.377 $\pm$ 0.015 \\
            \noalign{\smallskip}
            \texttt{[OIII]}    & & \textbf{5006.8}   & & 0.176 $\pm$ 0.020 & & 0.295 $\pm$ 0.010 & & 0.364 $\pm$ 0.015 \\
            \noalign{\smallskip}
            \texttt{MgI}       & & \textbf{5175.3}   & & 0.138 $\pm$ 0.019 & & 0.253 $\pm$ 0.010 & & 0.319 $\pm$ 0.014 \\
            \noalign{\smallskip}
            \texttt{NaD1}      & & \textbf{5889.950} & & 0.000 $\pm$ 0.017 & & 0.101 $\pm$ 0.008 & & 0.159 $\pm$ 0.013 \\
            \noalign{\smallskip}
            \texttt{NaD2}      & & \textbf{5895.924} & & $-$0.001 $\pm$ 0.017 & & 0.100 $\pm$ 0.008 & & 0.158 $\pm$ 0.013 \\
            \noalign{\smallskip}
            \texttt{NIIa}      & & \textbf{6549.86}  & & $-$0.101 $\pm$ 0.015 & & $-$0.010 $\pm$ 0.008 & & 0.042 $\pm$ 0.011 \\
            \noalign{\smallskip}
            \texttt{H$\alpha$}   & & \textbf{6562.8}   & & $-$0.103 $\pm$ 0.015 & & $-$0.012 $\pm$ 0.008 & & 0.040 $\pm$ 0.011 \\
            \noalign{\smallskip}
            \texttt{NIIb}      & & \textbf{6585.27}  & & $-$0.106 $\pm$ 0.015 & & $-$0.015 $\pm$ 0.008 & & 0.037 $\pm$ 0.011 \\
            \noalign{\smallskip}
            \texttt{S2}        & & \textbf{6716.0}   & & $-$0.123 $\pm$ 0.015 & & $-$0.035 $\pm$ 0.007 & & 0.016 $\pm$ 0.011 \\
            \noalign{\smallskip}
            \hline
         \end{tabular}
         \end{center}
\end{table*}

\section{Discussions and Outlook}
\subsection{Raman emissions and impact on astrophysical observations}
If an observation accidentally collides with the laser beam, depending on the wavelength range in which one is observing, it will be contaminated at either the laser line's wavelength, the secondary Raman emissions, or all of them. Near-infrared observations won't be affected by the collision, as the wavelength of the LGS and the O$_{2}$ and N$_{2}$ lines originating from Raman scattering are all in the visible range.\\
For spectroscopic observations, the impact of a collision with the laser beam depends on the spectrum of the observed object itself and the resolution of the spectra. At higher resolutions a contamination would have a different impact than at low resolutions, as it affects a narrower region of the spectrum. Also, if at the wavelength(s) of the laser line(s) the spectrum has only continuum, the superposed mission line will be easily recognizable, and can be either masked out, or the underlying continuum interpolated, or that wavelength region may be simply ignored in the analysis. Whether or how much an observation is affected by a collision with the laser beam, does not directly depend on the specific science case, but more on the observing mode in question, particularly the spectral resolution in case of a spectroscopic observation. A significant complication may occur when at the wavelength(s) of the laser line(s) the spectrum also exhibits spectral lines (being them emission or absorption lines).
These spectral lines may correspond to different lines at different redshifts.
Table \ref{TabRedshifts} shows the redshifts ranges for which several important spectral lines coincide with the LGS, the O$_{2}$ and N$_{2}$ vibrational lines and their respective rotational side-lines.
As one can see from the table, the redshift range in question is mainly for $z<1$. On a further consideration, these lines need to be subtracted, requiring a dedicated treatment and/or observing strategy (\citealt{weilbacher20}).\\
However, this would be potentially problematic for extended objects, when a second sky exposure is required to subtract the atmospheric lines, as in the case of IFU observations, or drift scan spectroscopy programs in the wavelength range compatible with the LGS, O$_{2}$ and N$_{2}$ rotational lines. In such cases, when the laser causes the emission of lines that vary in time, a good sky correction would not be achieved, leading the astronomer to assume that they are real lines, with the consequent creation of improbable models. Nevertheless, this should still be identifiable (i.e. as it is a more or less uniform emission) and the consequences would be contamination and partial degradation of the SNR of the scientific data.

\subsection{Impact of laser beam scattering}
It is always desirable to avoid collisions between scientific observations with visible-sensitive instruments and the laser beam. This implies an angular avoidance zone around the laser beam, during observations.\\
The scattering of the uplink laser beam consists of photons at 589 nm but also at O$_{2}$ and N$_{2}$ bands (see Table \ref{TabIntFlux}). They may contaminate the scientific observations. In case photon scattering affects the scientific observations, it does not necessarily make the observation useless, or diminish its quality. Notch filters or band-pass filters may mitigate the scattered light contamination in the sodium, O$_{2}$ and N$_{2}$ molecular bands, but retrofitting them in existing visible spectro-imaging instruments or telescope guiders could not always be a viable solution.\\
From Section \ref{laserphot} we know that during our experiment, imaging observations in the visible would have experienced an increase in the sky-background when the scientific FoV was at a distance <0.2 arcmin from the uplink beam, and <0.1 arcmin from the laser plume.\\
The LGS, O$_{2}$ and N$_{2}$ lines are located in the wavelength range of the Sloan R band, nevertheless other Sloan filters such as G, I, Z may be affected too by laser beam scattering. Sloan G may undertake the effects of long exposures due to a $\sim$2\% transmission at 589 nm, while Sloan I has Raman lines in its footprint (including the second vibrational levels from O$_{2}$ and O$_{2}$, CH$_{4}$, and H$_{2}$O), finally Sloan Z presents a transmission of $\sim$12\% at the location of the N$_{2}$($\nu_{2 \leftarrow 0}$) (\citealt{vogt17}). Likewise, Raman lines are also in the wavelength range of the Bessel R band. The Bessel V band might be affected by the LGS at 589 nm, but the transmission of a typical Bessel V filter is below 30\% at the wavelength in question (\citealt{rovilos09}), while all other filters won't be affected.

\subsection{Effects of the aerosols}\label{aerosol}
Aerosols in the atmosphere induce scattering of light through Mie scattering (\citealt{stra41}). For non-radially symmetric aerosols the scattering may occur in more complex ways (\citealt{dub06}; \citealt{gaug18}).\\
On the night of the 6th of July 2017, the GTC dust sensors measured negligible concentration of dust in the atmosphere (dust is a contributor to Mie scattering), with an average of 2.44 $\mu$g/m$^3$ for particles smaller than 25 $\mu$m, and 0.68 $\mu$g/m$^3$ for particles larger than 25 $\mu$m. The same happened for the night of the 7th of July 2017, when we experienced a negligible increase of dust concentration, with an average of 3.74 $\mu$g/m$^3$ for particles smaller than 25 $\mu$m, and 0.84 $\mu$g/m$^3$ for particles larger than 25 $\mu$m. The measured values can be used to calculate the total optical depth $\tau_\lambda$  as the sum of the contribution of the optical depth of each particle size (\citealt{lom11}):

\begin{equation}
\tau_\lambda = \sum_{\rho}\tau_\lambda (\rho) = \sum_{\rho} N(\rho)\sigma_\lambda (\rho)
\end{equation}
where $N(\rho)$ is the column density of particles of radius $\rho$, and $\sigma_\lambda (\rho)$ is the wavelength dependent absorption cross section from the Mie theory for particles of radius $\rho$. Making use of satellite measurements, the dust density over the Canary Islands has been found to be approximately constant at altitudes between 2500 and 5000 meters above sea level, dropping close to zero above 5000 m (\citealt{smirnov98}; \citealt{hsu99}; \citealt{alpert04}; \citealt{lom08}). Following this evidence, at the ORM the column densities can be approximated to the volume densities multiplied by 2500 m (\citealt{lom11}).\\
The calculated optical depths in R band for the nights of 6 and 7 July 2017 result to be $\leq$0.01, in excellent agreement with those in the visible from \citet{smirnov98} obtained for clean days from an altitude of 2356 meters above sea level in the nearby island of Tenerife (see Table 2 in the mentioned paper).\\
In this situation, additional effects induced by the presence of dust in the atmosphere (i.e. the Saharan dust called \textit{Calima}) cannot be evaluated in this Paper. It is widely demonstrated that the \textit{Calima} may have non-negligible effects on astrophysical observation at the ORM, for example by slightly increasing both the atmospheric extinction in visible wavelengths (\citealt{lom08}) and the thermal emissivity in the near-infrared (\citealt{lom11}). It has of course also an impact on the transmission and scattering of the LGS laser beam.\\
In such scenario, further observations of the LGS and its uplink will be crucial to evaluate the impact on the scattered light and the Raman emission in presence of dust in the atmosphere. We do intend to present a new proposal to repeat our observations under dusty conditions after the publication of this Paper. 

\subsection {Implications for the Laser Traffic Control System}
Laser Traffic Control Systems are algorithms designed to prevent the collision of the scientific FoV of one or more telescopes with the uplink beam of a propagating laser, or the LGS itself. The avoidance is achieved by calculating the geometric position of multiple telescopes FoV cones and the laser beam cylinder in the sky (\citealt{summ03}). A schematic representation of the geometry of a collision between the laser and the scientific FoV of a telescope can be found in Figure 2 in \citet{sant16}. The LTCS geometric calculations take into account:
\begin{enumerate} 
 \item telescope and laser coordinates and altitude above sea level;
 \item laser and telescope pointing coordinates (RA,DEC and equinox), thus their ALT/AZ and their dynamic variation during the targets tracking;
 \item sideral or non-sideral tracking;
 \item laser beam diameter;
 \item the angular separation between the telescope optical axis and the laser propagation direction;
 \item collecting surface, FoV, observational band and photon detection efficiency of the affected telescope.
\end{enumerate}
The outcome of the collision avoidance software provides the distance to the crossing point and its altitude above sea level, triggering the decision on the priority of one telescope over the other for a certain pointing direction, in the case of a low, medium or high probability of contamination of the laser scattered light in the FoV of telescopes - if observing at wavelengths bands affected by the propagating laser. Such contamination may impact the scientific operations in several ways (\citealt{gaug18}):
\begin{description}
    \item[$\bullet$] if active laser avoidance is chosen, the observing duty cycle will be affected;
    \item[$\bullet$] telescope's auto-guiding, active optics loops if any, and pointing precision may be mislead;
    \item[$\bullet$] false triggers may be generated for those facilities using triggering of image read-out (like Cherenkov telescopes).
\end{description}
Notch filters or band-pass filters are used to prevent some of the mentioned problems (\citealt{schal10}; \citealt{gaug18}), nevertheless this is not always possible. For the mentioned reasons, the collision avoidance may still be the preferred solution; not the least, if photometry contamination from Raman emissions generated by the laser uplink beam at wavelengths different from the laser are important for the scientific observations planned.\\
At the ORM, the code that the LTCS runs is the version 1.x from 2005. The actual version of the LTCS allows parameters configuration for every telescope, i.e. to set an appropriate minimum distance between the uplink laser beam and the "colliding" line-of-sight. In practice, different telescopes may set their own restrictions on the basis of their requirements. Nevertheless, at present this feature is not used, and a wide fixed value of 1.0 degree around the target (0.5 degrees in each direction) is assigned (private communication). At Paranal Observatory, a configuration value (dithering angle) can be established for each instrument on every telescope, nevertheless by default it is set to 20 arcmin around the target (10 arcmin in each direction), which shall cope for any offset the instrument may trigger without changing the telescope Guide Star (private communication).\\
As shown, different observatories could implement different LTCS settings, but at present those settings are the same for all telescopes involved in the same LTCS. To better optimize operations at crowded observatories, we believe that the existing feature that allow to assign a proper minimum distance for each telescope or instrument should be fully exploited. This can be achieved calculating the uplink defocus at every altitude, taking into account the observation geometry. The defocus will change with the elevation, so a mathematical function may be embedded in the LTCS code that periodically calculates it and launches a warning when a potential photometric contamination is detected in advance.\\ On the other hand, notch filters will avoid spectral contamination from the sodium line, but not from the Raman emission, so the proposed LTCS upgrades should also include a calculation for every altitude of the expected fluxes of (at least) the unresolved Q-branch of the O$_{2}$ and N$_{2}$ molecules first vibrational transition.\\
For our instrumental setup, we have shown in Section \ref{laserphot} that the 589 nm uplink beam scattered light in Sloan R band, for a telescope focused on the uplink beam up to 35 km, spreads up to $\pm$0.2 arcmin from the center of the beam. When focused at infinity during real astronomical observations, a telescope will already be affected at larger angular distances from the (out of focus) uplink laser beam. At which distance and with which scattered flux the observations are affected depends on the telescope optics, the telescope instrument and the observation geometry. Hence the minimum distance has to be scaled to the specific observing conditions taking into account the specific setup. \\
In the case of focusing the telescope at the LGS mesospheric plume, the scattering extends up to $\pm$0.1 arcmin and this can be considered a more definitive result, due to the negligible scattering in the mesosphere. The impact of the Raman lines emission, compared with the 589 nm scattered photons, is orders of magnitudes smaller than the primary scattering, as given in Table \ref{TabIntFlux}. Still, the presence of the Raman emission lines might be relevant in certain observation setups, and should be taken into account informing the users about their radiance vs altitude (scaling Figure \ref{fig:FigSter} according to the photon emission ratios measured in comparison with the LGS).\\
Our recommendation is that the LTCS software (at least for the GTC at ORM) should consider our findings. In the light of the performed observations and analysis, we find that the current LTCS practice of setting a fixed minimum angular distance value applied for all participating telescopes in an observatory is not sufficiently safe in terms of avoiding photometric contamination. More in general, our findings could be used to define the collision distance as a function of beam crossing altitude between the uplink laser beam and the colliding telescope pointing. That is, consider the minimum angular separation of the scientific FoV cone of the affected telescope from the laser propagating direction, depending on the altitude at which the collision may occur:
\begin{description}
    \item[$-$] at least $\pm$0.2 arcmin from the center of the beam for altitudes <50 km, that is where the atmospheric scattering becomes in practice negligible, using the measured radiance and scaling the received flux  to take into account the instrumental setup and defocus - which will require to scale the minimum collision distance accordingly;
 \vspace{0.1cm}
    \item[$-$] at least $\pm$0.1 arcmin from the center of the beam for collisions with the LGS plume when it is in focus, and otherwise scaled for the specific telescope, considering the defocus effects.
\end{description}
The LTCS algorithm should therefore take into consideration the effects of defocus for the various altitudes, taking as parameters as well the observing telescope focal length, diameter and of course recompute, or scale, the intercepted laser beam angular distance accordingly. This will improve the avoidance of scattered light and its contamination of the scientific FoV, as well as the occurrence of the LGS and the Raman lines photons in spectroscopic data. It will also optimize the set distance at which a beam crossing alarm is created, which in certain configurations may be smaller than the observatory-wide set beam crossing angular distance. The user should be accordingly warned about Raman bands emission and intensities, so the effect on his scientific instrument and procedures can be handled.\\
This is a recommendation for an update of the functionalities of the current LTCS software used at astronomical sites, based on the results obtained. 

\section*{Acknowledgements}
We thank the Referees for their useful comments and the GTC telescope operators and astronomers for their skilled and dedicated support with our very unusual observing mode requests.\\
We also thank Tim Morris and Matthew Townson at Durham University, and Philippe Duhoux at ESO for the useful information about the ORM and Paranal Observatory LTCS systems.\\
The experimental work has been done using the ESO Laser Guide Star Unit installed and supported at the ORM by the ESO Technology Development Program, Laser Systems R\&D work package 994. Data for this paper have been obtained with the 10.4m Gran Telescopio CANARIAS operated on the island of La Palma by the Grantecan S.A. at the ORM, under the International Time Programme of the CCI (International Scientific Committee of the Observatorios de Canarias of the IAC).

\section*{Data Availability}
The data underlying this article will be shared on reasonable request to the corresponding author.


   \vspace{-0.1cm}

\bibliographystyle{mnras}



\bsp	
\label{lastpage}
\end{document}